\documentclass[11pt]{article}

\usepackage{amsmath}
\usepackage{amssymb}
\usepackage{amsmath}
\usepackage{amssymb}
\usepackage{amscd}

 \newcommand{\beq}{\begin{equation}}
\newcommand{\eeq}{\end{equation}} 
\newcommand{\bea}{\begin{eqnarray}}
\newcommand{\eea}{\end{eqnarray}}

 \newcommand{\qm}{quantum
mechanics} 
\newcommand{\ca}{$C^*$-algebra} 
 \newcommand{\rep}{representation}

\newcommand{\Hs}{Hilbert space}

 \newcommand{\ovl}{\overline}
 \newcommand{\til}{\tilde}
\newcommand{\raw}{\rightarrow}

\newcommand{\ot}{\otimes} 
\newcommand{\la}{\langle} \newcommand{\ra}{\rangle}

\newcommand{\x}{\times} 
 
\newcommand{\Tr}{\mbox{\rm Tr}\,}

\newcommand{\half}{\mbox{\footnotesize $\frac{1}{2}$}}

\newcommand{\quar}{\mbox{\footnotesize $\frac{1}{4}$}}

\newcommand{\er}{\eqref}
\newcommand{\al}{\alpha} 
\newcommand{\gm}{\gamma} 
\newcommand{\dl}{\delta} \newcommand{\Dl}{\Delta}
 \newcommand{\varep}{\varepsilon}

\newcommand{\lm}{\lambda} 
\newcommand{\rh}{\rho} \newcommand{\sg}{\sigma}
 \newcommand{\ta}{\tau} 
 
 \newcommand{\ps}{\psi} \newcommand{\Ps}{\Psi}
\newcommand{\om}{\omega} 
\newcommand{\CA}{{\mathcal A}} \newcommand{\CB}{{\mathcal B}}

 \renewcommand{\H}{{\mathcal H}}

 \newcommand{\CS}{{\mathcal S}}

\newcommand{\C}{{\mathbb C}} 
 
\newcommand{\N}{{\mathbb N}} \newcommand{\R}{{\mathbb R}}
 \newcommand{\Z}{{\mathbb Z}}
  \makeatletter
\newskip\tempskip \def\endproof{{\parfillskip24\p@ plus\@ne
fil\@@par}\tempskip\prevdepth
\ifdim\lastskip=\z@\tempskip\z@\else\vskip-\lastskip
\ifdim\tempskip>4\p@ \tempskip.5\tempskip \else \tempskip\z@\fi\fi
\nobreak\vskip-\baselineskip\vskip-\tempskip\noindent\hbox
to\hsize{\hfill
$\blacksquare$}\par\vskip\tempskip\vskip\abovedisplayskip\@doendpe}
\makeatother \makeatletter
\newskip\tempskip \def\endiproof{{\parfillskip24\p@ plus\@ne
fil\@@par}\tempskip\prevdepth
\ifdim\lastskip=\z@\tempskip\z@\else\vskip-\lastskip
\ifdim\tempskip>4\p@ \tempskip.5\tempskip \else \tempskip\z@\fi\fi
\nobreak\vskip-\baselineskip\vskip-\tempskip\noindent\hbox
to\hsize{\hfill
$\Box$}\par\vskip\tempskip\vskip\abovedisplayskip\@doendpe}
\makeatother 

\renewcommand{\ca}{C*-algebra}
\newcommand{\mmp}{measurement problem}
\newcommand{\ssb}{{\sc ssb}}

\newcommand{\uaw}{\uparrow}
\newcommand{\dow}{\downarrow}

\topmargin = - 0.5 cm \textheight = 23 cm \textwidth = 15 cm
\begin{document} 
\pagenumbering{arabic} \setlength{\unitlength}{1cm}\cleardoublepage
\thispagestyle{empty}
\title{Spontaneous Symmetry Breaking in Quantum Systems:\\ Emergence or Reduction?\footnote{Final version, to be published in \emph{Studies in History and Philosophy of Modern Physics}.}}
\author{N.P. (Klaas)  Landsman\footnote{
 Institute for Mathematics, Astrophysics, and Particle Physics,
   Radboud University Nijmegen, Heyendaalseweg 135, 6525 {\sc aj} Nijmegen, the Netherlands.
Email: \texttt{landsman@math.ru.nl}}}
\date{\today}
\maketitle \begin{center} 
\emph{Dedicated to the memory of G\'{e}rard Emch (1936--2013)}
 \end{center}
 \bigskip
 
 \begin{abstract}
 \medskip
 
 Beginning with Anderson (1972), spontaneous symmetry breaking (\ssb) in infinite quantum systems is often put forward as an example of (asymptotic) emergence in physics, since in theory no finite system should display it. 
Even the correspondence between theory and reality is at stake here, since numerous real materials show \ssb\ in their ground states (or  equilibrium states at low temperature), although they are finite. 
Thus  against what is  sometimes called `Earman's Principle',
a genuine physical effect (viz.\ \ssb) seems theoretically recovered only in some  idealization (namely the thermodynamic limit), disappearing as soon as the  idealization is removed.

 We  review the well-known arguments that (at first sight) no finite system can exhibit \ssb, using the formalism of algebraic quantum theory in order to control the thermodynamic limit and unify the description of 
 finite- and infinite-volume systems.  Using the striking mathematical analogy between the  thermodynamic limit and the classical limit, we show that a similar situation obtains in quantum mechanics (which typically forbids \ssb) versus classical mechanics (which allows it).  
 
This discrepancy between formalism and reality is quite similar to the measurement problem (now regarded as an instance of asymptotic emergence), and hence
we address it in the same way, adapting an argument of the author and Reuvers (2013) that was originally intended to explain the collapse of the wave-function within conventional quantum mechanics. 
Namely, exponential sensitivity to (asymmetric) perturbations of the (symmetric) dynamics as the system size increases causes symmetry breaking already in finite but very large quantum systems. This provides continuity between finite- and infinite-volume descriptions of quantum systems featuring \ssb\ and hence restores Earman's Principle (at least in this particularly threatening case). 

\end{abstract}
\bigskip

\begin{small}
\begin{center}
\textbf{Motto} 
\end{center}
\begin{quote}
``The characteristic behaviour of the whole \emph{could} not, even in theory, be  
 deduced from the most complete knowledge of the behaviour of its  components, taken separately or in other combinations, and of their proportions and arrangements in this whole. This is what I understand by the `Theory of Emergence'. I cannot give a conclusive example of it, since it is a matter of controversy whether it                                                       
  actually applies to anything.''
  
  \hfill (Broad, 1925, p.\ 59)
\end{quote}
\end{small}

 \newpage
\section{Introduction}
\subsection{Emergence: from philosophy to physics}\label{EiP}
In a philosophical context, the notion of `emergence' is usually traced to J.S.\ Mill, who did not actually use this terminology himself but  drew attention to ``a distinction so radical, and of so much importance, as to require a chapter to itself.'' (wow!) 
The distinction in question is the one between the principle of the ``Composition of Causes'', according to which the joint effect of several causes is identical with the sum of their separate effects, and the negation of this principle. For example,  in the context of his overall materialism, Mill  believed that 
although all ``organised bodies'' are composed of material parts, \begin{quote}
``the phenomena of life,  which result from the juxtaposition of those parts in a certain manner, bear no analogy to any of the effects which would be produced by the  action of the component substances considered as mere physical agents.  To whatever degree we might imagine our knowledge of the properties of the several ingredients of a living body to be extended and perfected,  it is certain that no mere summing up of the separate actions of those elements will ever amount to the action of the living body itself."
 \hfill Mill (1952 [1843],  p.\ 243).\footnote{Quotation taken from O'Connor and Wong (2012).}
\end{quote}
This kind of thinking began what is now called `British Emergentism' (cf.\ Stephan, 1992;  McLaughlin, 2008; O'Connor and Wong, 2012), a school of thought which included A.\ Bain and G.H.\ Lewes in the 19th century and
virtually ended with C.D.\ Broad (who has our sympathy over Mill because of the doubt he expresses in our motto on the title page).
 Of this group, the most modern views seem to have been those of S.  Alexander, who, as paraphrased in O'Connor \& Wong (2012), was committed to a view of emergence as 
\begin{quote}
``the appearance of novel qualities and associated, high-level causal patterns which cannot be directly expressed in terms of the more fundamental entities and principles. But these patterns do not supplement, much less supersede, the fundamental interactions. Rather, they are macroscopic patterns running through those very microscopic interactions. Emergent qualities are something truly new (\ldots), but the world's fundamental dynamics remain unchanged.'' 
\end{quote} 
 Alexander's idea  that emergent qualities ``admit no explanation'' and had ``to be accepted with the `natural piety' of the investigator" seems to foreshadow the contemporary notion of explanatory
  emergence.\footnote{Seen as a branch of `epistemological' emergence. Philosophers   distinguish between \emph{ontological} and \emph{epistemological} reduction or emergence,   but  ontological emergence seems a relic from the days of vitalism and other immature understandings of physics and (bio)chemistry (including the formation of chemical compounds, which Broad and some of his contemporaries
still saw as an example of emergence in the strongest possible sense, i.e., falling outside the scope of the  laws of physics). Recent literature, including the present paper, is concerned with epistemological (or rather \emph{explanatory}) emergence.}  More precisely, the authorities wrote:
\begin{quote}
``The concept of \emph{emergence} has been used to characterize certain phenomena as `novel', and this not merely in the psychological sense of being unexpected, but in the theoretical sense of being unexplainable, or unpredictable, on the basis of information concerning the spatial parts or other constituents of the systems in which the phenomena occur, and which in this context are often referred to as `wholes'.'' \hfill  (Hempel \&\ Oppenheim, 2008 [1965], p.\ 62). 
\end{quote}
More recently, Silberstein (2002, p.\ 92)  states (paraphrased)  that a higher-level theory $\mathsf{H}$ 
\begin{quote}
``bears predictive/explanatory emergence with respect to some lower-level theory $\mathsf{L}$ 
 if $\mathsf{L}$ cannot \emph{replace} $\mathsf{H}$, if $\mathsf{H}$ cannot be \emph{derived} from $\mathsf{L}$ [i.e., $\mathsf{L}$ cannot reductively explain $\mathsf{H}$], or if $\mathsf{L}$ cannot be shown to be \emph{isomorphic} to $\mathsf{H}$." 
\end{quote}
In similar vein, Batterman (2002, p.\ 20) paraphrases Kim as stating that emergent properties are neither explainable nor predictable 
 even from ``exhaustive information concerning their basal [i.e., lower-level] conditions''.
 Wayne \& Arciszewski (2009, p.\ 852) propose that ``the failure of reductive explanation is constitutive of emergence in physics''. Last but not least,  the John Templeton Foundation defines (strongly)
 emergent phenomena as those which ``are, in principle, not derivable from the laws or organizing principles for, or even from an exhaustive knowledge about, their constituents''.\footnote{From \texttt{www.templeton.org/what-we-fund/funding-competitions/the-physics-of-emergence}, online at least from 2011--2013. The {\sc jtf} specifically funds research on emergence in this `strong' sense.} 

 In giving  quotations like this, we are interested not so much  in giving a precise definition of emergence (which may well be impossible and arguably also useless for such a broad concept), but rather in stressing that it is usually meant to be the very opposite  of the  idea of `reduction', or `mechanicism', as Broad (1925) calls it. 
Indeed, for many authors this  opposition seems to be  the principal attraction of emergence.
In principle, two rather different notions of reduction then  lead (contrapositively)  to two different kinds of  emergence, which are easily mixed up but should be distinguished (Norton, 2012):
 \begin{itemize}
\item the reduction of a whole (i.e., a composite system) to its parts;
\item the reduction of a 'higher-level' theory $\mathsf{H}$ (also known in physics as a 
`phenomenological' theory, or in philosophy as a `reduced' theory) 
to a lower-level one $\mathsf{L}$ (called a `fundamental' or a 'reducing' theory in physics and philosophy, respectively).
\end{itemize}
In older literature, e.g.\ concerned with the reduction of chemistry to physics (still challenged by Broad (1925)), the first notion also referred to `wholes' consisting of a \emph{small} number of particles. In this case, the possibility of whole-part emergence seems a lost cause in the sense originally intended,\footnote{ ``If a characteristic of a whole is counted as emergent simply if its occurrence cannot be inferred from a knowledge of all the properties of its parts, then (\ldots)  no whole can have any emergent characteristics. Thus (\ldots) the properties of hydrogen include that of forming, if suitably combined with oxygen, a compound which is liquid, transparent, etc.''
  (Hempel \&\ Oppenheim,  2008 [1965], p.\ 62).} but it might be revived in the context of entanglement. We will not go into this here, but instead follow
 Anderson (1972), who initiated the modern discussion on emergence  in physics in emphasizing  the possibility of emergence in \emph{large} systems.\footnote{Surprisingly, Anderson actually avoids the term `emergence' but speaks about new laws and even ``a whole new conceptual structure'' that each (higher) level can acquire. } In particular,   Anderson claimed  \ssb\ to be an example of emergence (if not \emph{the} example),  duly adding that one really had to take the $N\raw\infty$ limit. Thus 
the interesting case for emergence in the first (i.e.\ whole-part) sense  arises if the `whole' is strictly infinite, as in the thermodynamic limit of quantum statistical mechanics. The latter plays the role of the lower-level theory, and, as we shall see, there are two natural choices for the corresponding higher-level theory describing the `infinite whole', namely:
\begin{itemize}
\item  classical thermodynamics as a theory of \emph{macroscopic} (or \emph{global})  observables;
\item infinite-volume  quantum statistical mechanics as a theory of  \emph{local} observables. 
\end{itemize} 
  But such consideration take us straight into the realm of the second kind of emergence, especially if the 
inter-theory relation is asymptotic in the sense that the higher-level theory is a limiting case of the lower-level one. 

The conclusion, then, is that in the interesting case of whole-part reduction,  namely when the whole is infinite, one is actually dealing with asymptotic inter-theory reduction, so that the distinction between the two kinds of emergence is blurred. The other case we will consider, viz.\ the classical limit of quantum mechanics (as the lower-level theory, where the higher-level theory is classical mechanics) does not fall under the umbrella of the whole-part dichotomy and is purely a case of asymptotic inter-theory reduction.

Indeed, it is the inter-theoretic reduction perspective (rather than the whole-part notion of reduction) that will be the most useful for us, because we will display striking analogies between \ssb\ in classical mechanics (as a limiting theory) and \ssb\ in thermodynamics (\emph{idem}). In fact, it is exactly this `asymptotic' situation that contemporary research on emergence in physics is mostly concerned with. In particular, it has been claimed that in certain situations some properties of the higher-level theory (like \ssb) could in principle `emerge' asymptotically, in that \emph{nothing} in the lower-level theory corresponds or gives rise to those properties, not even if one is close to the limit. In this case, the higher-level theory is claimed to be `ineliminable', or, in other words, the lower-level theory is said to be `explanatorily insufficient'.

In order to make such claims one has  to assume that  the limiting higher-level theory  is well defined and  understood by itself (i.e., ignoring its nature as a limiting theory); 
historically it typically preceded 
the lower-level theory, as in the cases of classical thermodynamics vis-\`{a}-vis (quantum) statistical mechanics, and classical mechanics versus quantum mechanics. In contrast,  local quantum statistical mechanics in infinite volume became a well-defined theory in the 1970s only after the construction of its finite-volume counterpart, but
 we  still have  widely accepted and well-understood construction of it (Bratteli \&\ Robinson, 1997; Haag, 1992).

In summary, we speak of \emph{asymptotic emergence} if the following conditions are satisfied:\footnote{
Note that the lower-level theory $\mathsf{L}$ is really a family of theories, parametrized (in this case) by $N$ or $\hbar$.}
\begin{enumerate}
\item A higher-level theory $\mathsf{H}$ is a limiting case of some lower-level theory $\mathsf{L}$;
\item Theory $\mathsf{H}$ is  well defined and  understood by itself;
\item Theory $\mathsf{H}$ has `emergent' features that cannot be explained by $\mathsf{L}$.
\end{enumerate}

This scenario has been proposed and developed most notably by  Batterman (2002, 2011) and Rueger (2000, 2006), who gave a number of beautiful examples illustrating their point.\footnote{For criticism of a different kind from the present paper see  Hooker (2004),  Belot (2005) with a reply by Batterman (2005),  Wayne \& Arciszewski (2009), Butterfield (2011), and Menon \& Callender (2013).} In our examples, the theories $\mathsf{H}$ and $\mathsf{L}$ will be chosen and related as follows:
\begin{itemize}
\item $\mathsf{H}=$ quantum mechanics, taken to be the $\hbar\raw0 $ limit of $\mathsf{L}=$ quantum mechanics;
\item $\mathsf{H}=$ thermodynamics as the $N\raw\infty$ limit of $\mathsf{L}=$ quantum statistical mechanics;
\item $\mathsf{H}=$ local quantum statistical mechanics in infinite volume, seen as the $N\raw\infty$ limit of $\mathsf{L}=$ quantum statistical mechanics (defined for $N<\infty$, as also above).
\end{itemize}
\subsection{Two questions, two examples}\label{twosec}
A distinction  needs to be made in principle between two different questions about the possible relationship between theories $\mathsf{H}$ and $\mathsf{L}$, which may easily be confused in practice:
\begin{enumerate}
\item Can $\mathsf{H}$ (indeed) be derived as a limiting case of $\mathsf{L}$ (i.e., some family $\mathsf{L}_{\bullet}$, say $\mathsf{L}_N$)?
\item Do the  theories $\mathsf{L}_N$ for sufficiently large $N$ approximately behave like $\mathsf{H}$?
\end{enumerate}
At first sight these questions appear to be virtually identical, and indeed they are very closely related, but conceptually they are quite different:
the first is a question about the higher-level theory (and its possible origin in the lower-level one),
 whereas the second  concerns the  lower-level theory (in which the higher-level theory plays an ancillary role). For example, the  examples of the rainbow and the {\sc wkb} approximation in Batterman (2002) deal with the \emph{second} question, as does the research field of quantum chaos, whereas the author's (older) work on the relationship between classical and quantum mechanics tries to answer the \emph{first} (Landsman, 1998).\footnote{The review Landsman (2007) covers \emph{both}.}
 
The difference between these questions, as well as their sensitivity to  the notions of limit and approximation involved, may also be illustrated by Butterfield's (2011) sequence $(g_N)$ of functions $g_N:\R\raw\R$, defined by 
\begin{eqnarray}
g_N(x)&=&-1\:\:(x\leq -1/N);\\
g_N(x)&=&Nx\:\:  (-1/N\leq x\leq 1/N);\\
  g_N(x)&=&1\:\: (x\geq 1/N). 
\end{eqnarray}
Each $g_N$ is continuous, but
as $N\raw\infty$ this sequence converges pointwise to the discontinuous function $g_{\infty}$ given by 
\begin{eqnarray}
g_{\infty}(x)&=&-1\:\: (x<0);\\
  g_{\infty}(0)&=&0;\\
g_{\infty}(x)&=&1\:\: (x>0). 
\end{eqnarray}
\begin{itemize}
\item If we take the theories $\mathsf{L}_N$ to be  simply the functions $g_N$ themselves and the notions of limit and approximation are pointwise, then the answer to both questions 
is ``yes''.
 \item  If, on the other hand, we define a new sequence of theories  $\mathsf{L}'_N$ by saying that:
 \begin{enumerate}
\item   $\mathsf{L}'_N=0$ if $g_N$ is continuous;
\item   $\mathsf{L}'_N=1$ if it isn't,
\end{enumerate}
 then with the same notion of limit etc.\  the answer to the first question remains ``yes'', yet the second answer is ``no''.
\end{itemize}
In this case, a reductionist would prefer $\mathsf{L}$, whereas an emergentist would probably go for $\mathsf{L}'$. It could be argued, however, that the latter is being unreasonably strict here, since his/her predicate depends on the behaviour of some  function at just one point,\footnote{More generally,  by Lusin's theorem from measure theory the difference between continuity and discontinuity is just a matter of epsilonics; see e.g., Stein and Shakarchi (2005), p.\ 34. } and even within this limited scope, the discontinuity of the limit function $g_{\infty}$ at zero is well approximated by the derivatives $g_N'(0)=N$ converging to infinity.

Another instructive mathematical example, moving from single functions to function spaces but otherwise in the same spirit, is given by the following theory:\footnote{As we shall see (cf.\ \S\ref{cfcas}), the lower-level  theories $\mathsf{L}_N$  model certain primitive  aspects of a theory with $N$ degrees of freedom, of which the higher-level $\mathsf{H}$ is the `thermodynamic limit', which describes 
 a theory of quasi-local observables in the corresponding infinite system.}
\begin{enumerate}
\item  $\mathsf{L}_N=\ell(\underline{N})$ for $N<\infty$, where
$\underline{N}$ is the finite set $\{0,1,2,\ldots, N-1\}$ and $\ell(\underline{N})$ consists of all functions
$f:\underline{N}\raw\C$;
\item   $\mathsf{H}=\ell_0(\N)$, i.e., the space of all  functions $f:\N\raw\C$ that vanish at infinity (in other words, $\lim_{n\raw\infty} f(n)=0$), seen as a Banach space in the supremum-norm $\|\cdot\|_{\infty}$.\footnote{This norm is defined by $\|f\|_{\infty}=\sup\{\|f(n)\|,n\in\N\}$.}
\end{enumerate}
Also, define 
 maps $\iota_{M,N}:\ell(\underline{M})\raw \ell(\underline{N})\:\: (N\geq M)$ by 
 \begin{eqnarray}
 \iota_{M,N}(f)(n)&=&f(n)\:\: (n=0, \ldots, M-1);\\
 f(n)&=&0\:\: (n\geq M).
\end{eqnarray}
 Similarly, one has maps $\iota_{M,\infty}:\ell(\underline{M})\raw \ell_0(\N)$.
 
 We now answer the first question in the affirmative (whilst also making it precise for the case at hand), arguing
 that symbolically $\mathsf{H}=\lim_{N\raw\infty}\mathsf{L}_N$ in the following sense. For each $f\in\ell_0(\N)$ the sequence $(f_N)$, where $f_N\in\ell(\underline{N})$ is given by  $f_N=f_{|\underline{N}}$, has the property that $\iota_{N,\infty}(f_N)$ uniformly converges to $f$ as $N\raw\infty$, i.e., 
 \begin{equation}
\lim_{N\raw\infty}\|f-\iota_{N,\infty}(f_N)\|_{\infty}=0. \label{limitf}
\end{equation}

 The second question also has an affirmative  answer, which for later use we state in a rather more abstract way than strictly needed at this point. We say that a sequence of functions $(f_N)$, where once again $f_N\in\ell(\underline{N})$,\footnote{To avoid any confusion: this notation does \emph{not} imply that  $f_N=f_{|\underline{N}}$ for some $f\in\ell_0(\N)$.}
  is \emph{local} if there exists some $M\in\N$ such that $f_N=\iota_{M,N}(f_M)$ for all $N\geq M$
 (this implies $f_N(n)=0$ for all $N\geq M$ and $n\geq M$). If  $f\in\ell_0(\N)$ has finite support, then  the restrictions $f_N=f_{|\underline{N}}$  form a local sequence. But this is not necessarily the case 
 for arbitrary $f\in\ell_0(\N)$  (since the condition just given in brackets is not satisfied if $f$ has infinite support). Thus we say that a sequence  $(f_N)$ with $f_N\in\ell(\underline{N})$ is \emph{quasi-local} if for each $\varep>0$ there is some $M=M(\varep)$ and some local sequence $(f_N')$ such that $\|f_N-f_N'\|_{\infty}<\varep$ for all $N>M$. As intended, sequences like $(f_{|\underline{N}})$ are indeed quasi-local for any $f\in\ell_0(\N)$.
 Conversely, a quasi-local sequence $(f_N)$ has a limit $f\in\ell_0(\N)$, given pointwise by $f(n)=\lim_{N\raw\infty} f_N(n)$; an elementary $\varep/3$ argument shows that this limit exists and that the ensuing function $f$ satisfies \er{limitf}.
 
 The precise sense, then, in which the $\mathsf{L}_N$ approximately behave like $\mathsf{H}$, is that any $f\in\ell_0(\N)$ may be approximated in the sense of  \er{limitf} by the quasi-local sequence $(f_N)$ with $f_N=f_{|\underline{N}}$, and that such $f$ may be reconstructed from the approximating sequence.
 
An asymptotic emergentist could  challenge this reasoning by adding the following predicate to $\mathsf{H}$: any $f\in\ell_0(\N)$ must have infinite support. This condition is  `emergent' in the limit $N\raw\infty$, since none of the approximants $f_{|\underline{N}}$ satisfy it.
However, the latter do satisfy it `up to epsilon', since up to arbitrary precision (in the supremum-norm, which is the sharpest available) any $f\in\ell_0(\N)$ can be approximated by functions with finite support. Hence
 our emergentist should  admit that for all practical purposes his/her limiting theory is indistinguishable from $\mathsf{H}$ and hence is `almost' obtained from the $\mathsf{L}_N$ by reduction. 
\subsection{Two  principles}
In examples from physics one would naturally expect our two  questions to practically coincide, or at least, to have the same answers. To be specific, let us move to the  context of the type of models from solid state physics  considered by Anderson (1972) and others, in which spontaneous symmetry breakdown (\ssb) is the (alleged) `emergent' property.  In that context,  all systems in reality are finite, so that we regard mathematical models of infinite systems as approximations or idealizations of finite ones.\footnote{See Norton (2012) for a fine distinction between approximations and idealizations, which does not seem to matter here.} The above expectation then corresponds to  what Jones (2006) calls \emph{Earman's Principle}:\footnote{Jones in fact quotes a slightly different version of the same idea from Earman (2003).}
  \begin{quote}
``While idealizations are useful and, perhaps, even essential to progress in physics, a sound principle of interpretation would seem to be that no effect  can be counted as  a genuine physical effect if it disappears
when the idealizations are removed.'' \hspace{5cm} (Earman, 2004, p.\ 191)
\end{quote}
In Earman's wake,  \emph{Butterfield's Principle} is the claim that in this and similar situations, where it has been argued (by other authors) that certain properties emerge strictly in some idealization (and hence have no counterpart in any part of the lower-level theory),
\begin{quote}
``there is a weaker,
yet still vivid, novel and robust behaviour  [compared to Butterfield's own definition of emergence as `behaviour that is novel and robust relative to some comparison class', which removes the reduction-emergence opposition] that
occurs before we get to the limit, i.e. for finite $N$. And it is this weaker behaviour
which is physically real.'' \hfill (Butterfield, 2011)
\end{quote}
Both principles are undeniably reductionist in spirit, but  they just appear to be common sense and it would seem to be provocative to deny them. 
But the problem for reductionists, and hence a potential trump card for asymptotic emergentists, is that both principles appear to be violated in three  important examples, namely the ones we mentioned at the end of \S\ref{EiP}.
In all cases, the (alleged) emergent property is \ssb, as detailed below, but the first two may also be considered in the light of the \mmp\ (see below).  Moreover, in these examples the apparent violation of Earman's and Butterfield's Principles is not a matter of mathematical epsilonics or philosophical hairsplitting, but  occurs rather dramatically, by a wide margin, and for essentially the same reason.

Namely, referring for concreteness's sake to the three examples at hand, both \ssb\ and the \mmp\ appear to pose what we  call the \emph{asymptotic emergence paradox}: 
\begin{itemize}
\item Reality, in which measurements have outcomes and symmetries can be broken,  approximately behaves like the higher-level theory $\mathsf{H}$, in which $N=\infty$ or $\hbar=0$;
\item Since actually $N<\infty$ and $\hbar<0$,  reality should be described by the lower-level theory $\mathsf{L}$ (in some asymptotic regime, where $N$ is very large and/or $\hbar$ is very small);
\item But whatever its parameter values $N$ or $\hbar$, $\mathsf{L}$ fails to describe measurements and \ssb;
\item So there is not only a lack of continuity  between $\mathsf{H}$ and $\mathsf{L}$, but even
a complete mismatch between reality and the theory $\mathsf{L}$ supposed to describe it.
\end{itemize}

We will now specify this in some more detail for the two cases of interest for us, which in turn are the measurement problem (briefly) and \ssb, our focus being the latter.
\subsection{Intermezzo: the \mmp\ revisited}
To see the \mmp\ of \qm\ as a special case of the (alleged) phenomenon of asymptotic emergence, one should \emph{define} measurement in the first place. We consider measurement undefined within \qm\ and hence have to construe it externally. 
 To do so, we follow Bohr (and experimental practice) in defining measurement as the establishment of a correlation between a quantum object and some other device that is necessarily quantum mechanical by nature as well, but which serves its purpose as a measurement apparatus  only if it is \emph{perceived} and hence \emph{described} classically (we will see in  detail how this is to be done in our examples). Briefly, measurement comes down to  `looking  at the quantum object through classical glasses' (Landsman, 2007, \S 3.1).
 
 Now Bohr seems to have taken it as a raw fact that measurements have outcomes, in that the post-measurement state of the measurement device (which by the above definition is necessarily described as a classical state) is \emph{pure} (or, in other words, is dispersion-free). But the whole problem  is that, at least at first sight, in many cases this does \emph{not} seem the be a prediction of \qm\ in its classical limit. Indeed, as we shall see in our examples, according to naive theory  typical post-measurement states are \emph{mixed}, being limits of Schr\"{o}dinger Cat states of the apparatus described quantum-mechanically.

In the context of asymptotic emergence,  the lower-level theory $\mathsf{L}$ is quantum mechanics or quantum statistical mechanics (of a finite system, that is), whereas the higher-level theory $\mathsf{H}$ is classical mechanics or thermodynamics, respectively. We never get tired of quoting Landau \&\ Lifshitz (1977, p.\ 3), asymptotic emergentists \emph{avant la lettre}:
\begin{quote}
 ``Thus \qm\ occupies a very unusual place among physical theories: it contains classical mechanics as a limiting case, yet at the same time it requires this limiting case for its own formulation.'' \end{quote}
 
So in this case, the `emergent' property of  $\mathsf{H}$ that (allegedly) fails to be induced by $\mathsf{L}$ is the purity of its states, or, more conceptually, the fact that there are facts (in the sense of sharply defined properties). 
But, in contrast to the case of \ssb\ to be discussed next, this immediately has to be qualified by saying that $\mathsf{L}$ contains many states that do converge to pure states of  $\mathsf{H}$ (such as coherent states): sometimes quantum theory does induce facts in its classical limit. Nonetheless, the \mmp\ is that, in measurement situations as defined above, states of $\mathsf{L}$ that according to experiment (where measurements have outcomes) should induce pure states on $\mathsf{H}$, at least in (naive) theory  fail to do so. 

Another qualification is that the world is really quantum mechanical (or so we think): the classical description of the measurement device called for by Bohr can never really be successful. But unlike the previous qualification, this one goes to the heart of the matter: if the \mmp\ were to remain unsolved, it would lead to the asymptotic emergence paradox describe at the end of the previous subsection. 

In sum, the \mmp\ of quantum theory is the following contradiction:
\begin{itemize}
\item The world behaves according to quantum theory, i.e., our lower-level theory $\mathsf{L}$;
\item Post-measurement states are \emph{described}  classically  by our higher-level theory $\mathsf{H}$;
\item According to observation, post-measurement states of $\mathsf{H}$ are \emph{pure};
\item Classical physics ($\mathsf{H}$) should be some limit ($N\raw\infty$, $\hbar\raw 0$) of quantum theory ($\mathsf{L}$); 
\item Yet in this limit, (pure) Schr\"{o}dinger Cat states of $\mathsf{L}$ converge to \emph{mixed} states of $\mathsf{H}$.
\end{itemize}
 \subsection{Spontaneous symmetry breaking}
We now turn to our main topic, which is  spontaneous symmetry breakdown  in large systems. Here the emergentist case seems much stronger in quantum physics than in classical physics, which in the context of phase transitions has  been adequately dealt with (Butterfield \&\ Bouatta,  2011; Melon \&\ Callender, 2013). In quantum theory, the crucial point is that for finite systems the ground state (or the equilibrium state at sufficiently low temperature) of almost any physically relevant Hamiltonian is unique and hence invariant under whatever symmetry group $G$ it may have. Hence, mathematically speaking, the possibility of \ssb, in the sense of having a family of ground states (etc.) related by the action of $G$, seems to be reserved for infinite systems (for which the arguments proving uniqueness  break down). 
 This leads to two closely related puzzles, one  devastating to the formalism, the other casting serious doubt on the link between theory and reality:
 \begin{itemize}
\item   The formal problem  is the  discontinuity between a large  system and a strictly infinite one: with appropriate dynamics, the former has a unique and invariant ground state, whereas the latter has a family of ground states, each asymmetric. Of course, 
 this is exactly the place where the asymptotic emergentist scores its goal in claiming that
 the thermodynamic limit is `singular' and hence
\ssb\ is an `emergent' property.
\item  The physics problem seems even more serious: nature displays \ssb\  in finite samples, yet the theory is unable to reproduce this and  seems to need the infinite idealization.
\end{itemize}
 Thus at least in case of \ssb, 
 quantum theory appears to violate Earman's and Butterfield's Principles. In other words, any description of a  large quantum system (i.e., large enough so that its physical manifestation displays \ssb) that models such a system as a finite system (which it is!) is empirically false (at least as far as \ssb\ is concerned), whereas a description that models it as an infinite system (which it is not!) is empirically adequate for \ssb. 
 
The situation involving the classical limit $\hbar\raw 0$ of \qm\ is analogous  (Landsman \& Reuvers, 2013). Take a particle moving in some $G$-invariant
potential whose absolute minima form a nontrivial $G$-space (in particular, each minimum fails to be $G$-invariant); the simplest example is
the symmetric double well  in dimension one, where $G=\Z_2$. 
For any $\hbar>0$ the quantum theory typically has a unique ground state,  peaked above the family of classical minima. In the limit $\hbar\raw 0$
the probability  distribution on phase space canonically defined by this state 
 does not converge to any one of the classical ground states, but converges to a symmetric convex sum or integral thereof. Nature, however, displays \emph{one} of the localized classical states. Yet for any positive $\hbar$, however small, the quantum ground state is delocalized  and shows no tendency towards one peak or the other. For $G=\Z_2$ this is the Schr\"{o}dinger Cat problem in disguise, which therefore seems to block any asymptotic derivation of classical physics from quantum physics. 
 
 A less familiar `intermediate' case interpolates between local quantum statistical mechanics in infinite volume and the classical limit of quantum mechanics. To wit, there are two very different different ways of taking the thermodynamic limit as far as the choice of observables of the corresponding infinite system is concerned (Landsman, 2007, Ch.\ 6):
 \begin{itemize}
\item the \emph{local} observables describe finite (but arbitrarily large) subsystems;
\item the \emph{global} (or \emph{macroscopic}) observables describe thermodynamic averages.
\end{itemize}
The former  retain the quantum-mechanical (i.e., noncommutative) character of the lower-level theory, whereas the latter form a commutative algebra, so that the higher-level theory acting as the limit of quantum statistical mechanics is classical. 
As we shall see, in this case the problem of \ssb\ is virtually the same as in the previous two.

Summarizing these three cases, the fundamental problem of \ssb\ is as follows:
\begin{enumerate}
\item One has a lower-level theory formulated in terms of a parameter $N < \infty$ (quantum statistical mechanics) or $\hbar > 0$ (quantum mechanics); 
\item The ground state (or  equilibrium state) of some Hamiltonian with  symmetry group $G$ is unique and hence $G$-invariant for any value of
$N < \infty$  or  $\hbar > 0$;
\item The  $N\raw  \infty$ or $\hbar \raw 0$ limit of the ground state (etc.) exists, is still $G$-invariant, but is now mixed (non-extremal i.e. not a pure thermodynamic phase);
\item The limit theories at $N=\infty$ (being either infinite-volume quantum statistical mechanics or classical thermodynamics) or $\hbar= 0$ (classical mechanics) exist  on their own terms (i.e. without taking any limit) and are completely understood;
\item These limit theories may display \ssb\ (depending on the model): they may have a family of 
$G$-variant pure ground states (extremal {\sc kms} states), forming a $G$-space;
\item  Nature may display \ssb, in which case physical samples modelled by such Hamiltonians behave like the limit theory 
(although in reality $N < \infty$  or  $\hbar > 0$);
\item Thus for any  $N < \infty$  or  $\hbar > 0$, the  theory neither approximates the limit theory nor models 
reality correctly: indeed, it spectacularly fails to do so! 
\end{enumerate}

These and similar problems with the thermodynamic limit  have been noted, though perhaps not exactly in those terms. But  as far as we know they have by no means been resolved.\footnote{The problem of \ssb\ in the thermodynamic limit is rarely if ever described 
 in conjunction with the analogous situation in the classical limit, not even in Landsman (2007), to which this paper is a successor.} For example,  in response to Earman,   Liu and Emch (2005, p.\ 155) first write that it is a mistake to regard idealizations as acts of ``\emph{neglecting the negligible}'', which already appears to deny both Earman's and Butterfield's Principles, and continue by:
\begin{quote}
``The broken symmetry
in question is \emph{not reducible} to the configurations of the microscopic parts of any \emph{finite} 
systems; but it should \emph{supervene} on them in the sense that for any two systems that 
have the exactly (sic) duplicates of parts and configurations, both will have the same spontaneous symmetry breaking  
in them because both will behave identically in the limit. In other words, the result of 
the macroscopic limit is determined by the non-relational properties of parts of the 
finite system in question.''
\hfill   Liu \& Emch (2005, p.\ 156)
\end{quote}  
 With due respect to especially our posthumous dedicatee, who was a master of mathematical  aspects of infinite-volume idealizations, it would be hard to explain even to philosophers (not to speak of physicists) how these comments solve the problem. Also in reply to Earman (2004), Ruetsche (2011) proposes to revise Earman's Principle as follows:\footnote{
As Ruetsche's notes herself, her Principle ``has the pragmatic shortcoming that we can't apply it until we know what (all) successors to our present theories are." (Ruetsche 2011, p.\ 336). A more pragmatic suggestion she makes, which is by no means inconsistent with her revision of Earman's Principle, is to realize that even quantum statistical mechanics in finite volume has an infinite number of degrees of freedom, as the underlying theory should ultimately be quantum field theory.}
 \begin{quote}
No effect predicted by a non-final theory can be counted as a genuine physical effect if it disappears from that theory's successors." \hfill  Ruetsche (2011, p.\ 336)
\end{quote}
Fortunately, both of these defensive manoeuvre  are unnecessary.
 \subsection{Saving the principles}\label{savingp}
 Namely, using the same idea in all cases, we shall establish the conceptual and mathematical continuity of \ssb\ in both
  the thermodynamic limit and the classical limit, and hence rescue Earman's and Butterfield's Principles. This idea will  be presented in detail
  in some models with a $\Z_2$-symmetry, viz.\ the  \emph{quantum Ising chain} in the thermodynamic limit 
  described using local observables, the closely related \emph{quantum Curie--Weisz model} in the same limit but now using global observables, and thirdly,  the quantum particle moving in a \emph{symmetric double-well} potential  in the classical limit. Although these models are somewhat special (we have adopted them because almost everything is known about them),  the conceptual lesson from these examples is clearly general, whilst
   the generalization to different models and larger symmetry groups seems to be a matter of technique.\footnote{
  Thus we prophylactically plead not guilty to the `case-study gambit' warned against by Butterfield (2011): 
``trying to support a general conclusion by
describing examples that have the required features, though in fact the examples are not typical, so that the attempt fails, i.e. the general conclusion, that all or most examples have the features, does not follow.'' Our examples are the `hydrogen atom(s)' of \ssb.}
      
Leaving the mathematical details to the main body of this paper,
 we now sketch this idea, which in the restricted context of the classical limit of quantum mechanics has already been proposed  as a possible solution to the measurement problem (Landsman \&\ Reuvers, 2013).
 Our present work strengthens this proposal, which we now extend from  the traditional classical limit  $\hbar\raw0$ to the thermodynamic limit $N\raw\infty$ and its associated classical realm.
The following scenario combines the technical work of Jona-Lasinio, Martinelli, \& Scoppola (1981a,b) and 
 Koma \& Tasaki (1994) on
 the classical limit and the thermodynamic limit, respectively,  also borrowing from Anderson (1952, 1984).
  
For any $N<\infty$,  let   $\Psi^{(0)}_N$ be the unique and hence $\Z_2$-invariant ground state of the quantum Ising chain with $N$ sites. Seen as a wave-function over spin configuration space, for large $N$  this is a Schr\"{o}dinger Cat state, doubly peaked above `all spins up' and `all spins down'. It 
 defines an algebraic state  $\psi^{(0)}_N:\CB_N\raw\C$ by taking expectation values, where 
 $\CB_N=\ot^N M_2(\C)$ (i.e.,  the $N$-fold tensor product of the $2\x 2$ matrices)
is the pertinent algebra of observables.
For $N\raw\infty$, there is a satisfactory notion of convergence of the sequence $(\psi^{(0)}_N)_N$ to a state $\psi^{(0)}_{\infty}$ on $\CB^l_{\infty}$, the algebra of (quasi-) local observables of the corresponding infinite system.\footnote{The 
quasi-local observables are local `up to $\varep$', see Landsman (2007, \S 6.2) or \S \ref{cfcas} below.}  
 The key point is that by $\Z_2$-invariance one obtains  
  \begin{equation}
\psi^{(0)}_{\infty}=\half (\psi^{+}_{\infty}+\psi^{-}_{\infty}),\label{Cat}
\end{equation}
i.e., a symmetric mixture of the two degenerate ground states of the limiting dynamics in infinite volume 
$\psi^{+}_{\infty}$ and $\psi^{-}_{\infty}$ (having all spins up and down, respectively). 

The situation for the Curie--Weisz model is analogous: the local algebras $\CB_N$ are the same as for the quantum Ising chain, but the  algebra of global observables $\CB^g_{\infty}$ is now commutative: it is isomorphic to the algebra $C(\mathsf{B}^3)$ of continuous functions on the three-ball $\mathsf{B}^3$ in $\R^3$, playing the role of the state space of a quantum-mechanical two-level system, so that its boundary is the familiar Bloch sphere. Once again, the quantum-mechanical ground state for finite $N$ is unique, and its limit as $N\raw\infty$ takes the form \er{Cat}, too. In this case the states $\psi^{\pm}_{\infty}$ are  points on the Bloch sphere, reinterpreted as probability measures on $\mathsf{B}^3$, and the left-hand side is their symmetric convex sum, seen as a  probability measure on the same space. Thus the state  \er{Cat} is again mixed.

Similarly, for $\hbar>0$,  the unique and hence $\Z_2$-invariant ground state $\Psi^{(0)}_{\hbar}$ of the double-well potential (in one dimension) defines an algebraic state
$\psi^{(0)}_{\hbar}:\CA_{\hbar}\raw\C$, where  $\CA_{\hbar}\cong \CA_1=K(L^2(\R))$ is the algebra of compact operators on the Hilbert space $L^2(\R)$ of square-integrable wave-functions, independent of $\hbar>0$. The family $(\psi^{(0)}_{\hbar})_{\hbar}$ has a  limit $\psi^{(0)}_0$ as $\hbar\raw 0$, which  is 
a state on the classical algebra $\CA_0=C_0(\R^{2})$ of continuous functions on phase space 
(taken to vanish at infinity for simplicity),\footnote{ In principle there would be a similar  ambiguity about the choice of the algebra of observables in the $\hbar\raw 0$ limit. We  here take the (quasi-)local observables, but in this case the delocalized observables  $C_b(\R^2)/C_0(\R^2)$ would be irrelevant, because unlike the previous case the \ssb\ problem here  is local.} i.e., a probability measure on phase space. Once again, this limit is the (by now) familiar symmetric convex combination 
\begin{equation}
\psi^{(0)}_0=\half (\psi^{+}_0+\psi^{-}_0),\label{Cat2}
\end{equation}
 where the Dirac (or `point') measure $\psi^{(\pm)}_0$ is concentrated at $(p=0,q=\pm a)$.\footnote{Here $\pm a$ are the the minima  of the double-well potential.}

The key idea to get rid of the \emph{mixed} limits \er{Cat} and \er{Cat2} in $\mathsf{H}$ of \emph{pure} Schr\"{o}dinger Cat states in $\mathsf{L}_{\bullet}$ 
is to go beyond the ground state and
also take the first excited state $\Psi^{(1)}_{\bullet}$ into account, where $\bullet$ is either $N$ or $\hbar$ as appropriate.\footnote{
For continuous symmetries, in which case one has an infinite number of low-lying states, this key idea may already be found in Anderson (1952), which influenced later authors like Koma \&\ Tasaki (1994).
}
 The essential point is that in our models, the energy difference $\Dl E_{\bullet}= E^{(1)}_{\bullet}- E^{(0)}_{\bullet}$ between $\Psi^{(1)}_{\bullet}$ and $\Psi^{(0)}_{\bullet}$ vanishes exponentially as $\Dl E_N\sim\exp(-C\cdot N)$
 for $N\raw\infty$,  or as
$\Dl E_{\hbar}\sim\exp(-C'/\hbar)$ for $\hbar\raw 0$, respectively. This means that asymptotically any linear combination of 
$\Psi^{(0)}_{\bullet}$ and $\Psi^{(1)}_{\bullet}$ is `almost' an energy eigenstate. For example, 
 consider the two linear combinations 
\begin{equation}
\Psi^{\pm}_{\bullet}=(\Psi^{(0)}_{\bullet}\pm \Psi^{(1)}_{\bullet})/\sqrt{2}. \label{wrong}
\end{equation}
These have the special virtue that  in the pertinent limit $N\raw\infty$ or $\hbar\raw 0$ the associated family of algebraic states $\psi^{\pm}_{\bullet}$ (weakly) converges to a \emph{pure} ground state $\psi^{\pm}_{\infty}$ (or $\psi^{\pm}_{0}$)
of the higher-level  theory, as opposed to the (symmetric) \emph{mixture} \er{Cat} or \er{Cat2} of such states.

Now the exponential decay of $\Dl E_{\bullet}$ implies that
almost any asymmetric perturbation (that does not vanish as quickly in the limit as $\Dl E_{\bullet}$, e.g., by being independent of $N$ or $\hbar$ or having at most a power-law decay) will eventually destabilize the ground state. Moreover, it does  so ever more effectively as $N$ grows or $\hbar$ decreases  (in other words, as the higher-level theory is approached). 
The  asymptotic behavior of the perturbed ground state $\Psi'_{\bullet}$ can be computed  using the `interaction matrix'  of Helffer and 
Sj\"{o}strand (1986),\footnote{See also Helffer (1988) and Simon (1985).
 This formalism was originally developed for Schr\"{o}dinger operators, but it  applies more generally; cf.\ \S\ref{TL} below.} 
 which reduces the  computation in question
to a $2\times 2$ matrix problem.
 Lo and behold:  depending on  the details of the perturbation,
$\Psi'_{\bullet}$ turns out to have the same
 asymptotic behavior as either $\Psi^+_{\bullet}$ or $\Psi^-_{\bullet}$, and hence converges to a \emph{pure} state $\psi^{\pm}_{\infty}$ (or $\psi^{\pm}_{0}$).
 
At the same time, the stability properties of the (pure) ground states of the limit theory guarantee
that the perturbed ground states of the lower-level theory are asymptotically insensitive to small perturbations.\footnote{In the context of the \mmp\ this means that measurements that already had outcomes (i.e.\ `dead' or `alive') from the beginning retain these outcomes, absolutely so in the limiting classical  theory and at least approximately so in reality; cf.\ the next two subsections.} Indeed, it is this very combination of the \emph{instability} of Schr\"{o}dinger Cat states of the lower-level theory with the \emph{stability} of pure ground states of the higher-level theory that, at least in principle, solves the problem of \ssb.\footnote{As well as the \mmp.  In the  context, see also Narnhofer \&\ Thirring (1996, 1999) for similar arguments (though adhering to the Copenhagen Interpretation with its associated incomprehensible `collapse of the wave-function').
There is no contradiction between this scenario and the linearity of the  Schr\"{o}dinger equation, because the algebraic states are quadratic in the state vectors, in combination with the limiting operations in $N$ or $\hbar$.
Since confusion around this point has arisen, we paraphrase footnote 30 in 
Landsman \&\ Reuvers (2013) here: the $2\times 2$ matrix approximation is solely  meant to  compute the asymptotic behavior of the perturbed ground state and hence to  illustrate the instability of Schr\"{o}dinger Cat states.
However, this approximation comes at the cost of 
 obscuring the stability of the pure ground states of the higher-level theory. We are indebted to Gijs Leegwater for discussions on this issue.\label{fwarning}
}
\subsection{Discussion}
For mathematics, our mechanism implies that the limit states are continuously approached.\footnote{We will make  this precise using the formalism of continuous fields of C*-algebras (Dixmier, 1977)
and  continuous fields of states thereon (Landsman, 2007, \S\S 4.3,5.1). }
For physics, it means  that the behaviour of real materials is well approximated by the lower-level theory at very large $N$ or very small $\hbar$. Our main claim in this respect has been that in this regime one should not look at the `official' ground state, which is unstable, but rather at some stable version to it, which already breaks the symmetry for finite $N$ or positive $\hbar$. However, this comes at a price: strictly speaking, such symmetry-breaking states in the lower-level theory are slightly unstable themselves (as opposed to their limits in the corresponding idealized higher-level theory). Indeed, for $N<\infty$ or $\hbar>0$, the symmetry-breaking states $\Psi^{\pm}_{\bullet}$ or  $\Psi^{(\pm)}_{\bullet}$
are unstable even for the unperturbed Hamiltonian, but for large $N$ or small $\hbar$ they move very slowly. For example, in our models with $\Z_2$-symmetry the transition time from $\Psi^{+}_{\bullet}$ to $\Psi^{-}_{\bullet}$ is approximately $2\pi/\Delta_{\bullet}$, which goes like $\exp(+CN)$ or $\exp(+C'/\hbar)$. This may seem enormous to theorists, who tend to think in terms of $N\sim 10^{23}$, but in the lab $N=10$ for a spin chain is already a large number!\footnote{Corresponding estimates for the Lieb--Mattis model of antiferromagnetism,
in which the ground state  breaks a continuous $SU(2)$ symmetry, are given in
van Wezel, van den Brink and Zaanen (2005). The work of 
van Wezel (2007, 2008, 2010), which was brought to our attention only after the first version of this paper appeared on the arXiv,  analyzes the spontaneous breakdown of continuous symmetries in a way compatible with (and indeed predating)
our treatment the discrete case, including the fundamental observation about the instability of the symmetric  ground state of a finite system under `infinitesimal' asymmetric perturbations.  Van Wezel (2010) also discusses 
 the connection between \ssb\ and the \mmp, though his proposed solution to this problem is quite different from ours. }

Either way,  the job is done by the exponential instability of the ground state (etc.) under asymmetric perturbations for large $N$ or small  $\hbar$, which should cause the system to pick a specific symmetry-breaking   state $\Psi^{\pm}_{\bullet}$ already for some \emph{finite} value of $N$ or \emph{positive} value of $\hbar$ (as opposed to their physically irrelevant limiting values $\infty$ or $0$).

Such perturbations may be induced either by the environment of the system, as in the `decoherence' (non) solution to the measurement problem (see Landsman \& Reuvers (2013) for more on this), or possibly by material defects.  In this sense, \ssb\  \emph{does not exist}:  for small systems reside in their naive symmetric ground state, large systems break symmetry explicitly (i.e., through the Hamiltonian), and infinite systems are not there. 

By the same token, in the real world  totally isolated finite  quantum systems (with perfectly symmetric Hamiltonians) should display \ssb, as the naive theory indeed predicts. In other words, truly macroscopic Schr\"{o}dinger Cat states are only possible if asymmetric perturbations can be totally suppressed (so in theory Schr\"{o}dinger's Cat can indeed exist!).

Where does this leave us? What seems dubious now about the idea of emergence (at least in the case at hand of \ssb) is, in our view, its opposition to reduction. So although for many this juxtaposition was the essence of the idea, it would seem wise to give it up. Thus we largely side with Butterfield (2011) in rescuing the side of emergence that talks about `novel and robust behaviour', but taking this novelty (with reference to our earlier quotation of Hempel and Oppenheim)
solely  in the psychological sense of being unexpected, rather than in the theoretical sense of being unexplainable or unpredictable.

Even so, one aspect of the idea that the higher-level theory is `ineliminable'  survives, namely the particular choice of observables
within the lower-level theory that guarantees the correct limiting behaviour. In quantum statistical mechanics, it is either the local or the macroscopic observables, each with their very peculiar $N$-dependence, that survive the limit $N\raw\infty$. Analogously, in quantum mechanics it is the semiclassical observables with their very peculiar $\hbar$-dependence that survive the limit $h\raw0$. 

On the one hand, the choice of these observables is dictated by the limit theories \emph{taken by themselves} (which therefore have to be known in advance) and is \emph{not} intrinsic to the underlying  lower-level theories. 
On the other hand, these specific observables are defined \emph{within} the latter, whose (mathematical) structure gives rise to the very possibility of singling them out. In other words,  it is in the nature of (quantum) statistical mechanics that one is able to define the small family of observables that in the thermodynamic limit behave according to the laws of thermodynamics, and it is quantum mechanics itself that allows the selection of the observables with the `right' $\hbar$-dependence (such as the usual position operator $x$ and the momentum operator $-i\hbar d/dx$).\footnote{Or,   more generally, the quantizations $Q_{\hbar}(f)$ of functions $f$ on classical phase space (Landsman, 2007).}
Hence in our view it merely seems a semantic issue whether the act of picking precisely the observables with appropriate limiting behaviour (or just the possibility thereof, or even merely their existence) falls within the scope of reduction, or instead is the hallmark of emergence. Either way, novel and robust behaviour is predicated on this act of choice. 

In the remainder of this paper we explain the technical details of this scenario.\footnote{Given the complexity of the underlying material from mathematical physics and the spatial constraints of a journal paper, our treatment cannot possibly be self-contained, however. For \emph{really} technical details or sub-details we briefly summarize the relevant arguments in footnotes, including  references to the literature for those who wish to study the background (or check our reasoning). We hope to give a full account in a planned textbook on the foundations of \qm\ called \emph{Bohrification}, scheduled for 2015. }
 Section \ref{Smodels} gives a description of our models and their symmetries. Section \ref{cf} describes some remarkable continuity properties of these models as $\hbar\raw 0$ or $N\raw \infty$, putting the folklore that such limits are `singular' in perspective.\footnote{This section is quite technical and may be skipped at a first reading.}
In section \ref{GS} we describe the ground states of our models, proving their discontinuity at $\hbar=0$ or $N=\infty$. But in the final section \ref{TL} we put the record straight by invoking the first excited states, which save the day and restore Earman's Principle and Butterfield's Principle, as outlined above. 

Finally, it will be obvious (at least to experts) that although we concentrate on ground states in this paper, the same scenario applies even more easily to equilibrium states at low temperature.\footnote{At high temperatures $T$ the issue of \ssb\  does not arise, since equilibrium states at given $T$ will typically be unique in both the lower- and  the higher-level theory (Bratteli \&\ Robinson, 1997; Haag, 1992).} Indeed, if  equilibrium states are formalized as {\sc kms} states (as usual in mathematical physics), then the role of pure ground states is played by \emph{primary} {\sc kms} states. For finite systems the former enjoy even better uniqueness properties than ground states, whilst for infinite systems the latter have even better stability properties. 
\subsection*{Acknowledgement}
The author is indebted to Rob Batterman for sending him the book (Batterman, 2002)  a decade ago, which triggered  his interest in Emergence as a topic amenable to mathematical analysis. This interest was subsequently kept alive by Jeremy Butterfield. Thanks also to Aernout van Enter, Bruno Nachtergaele, and Hal Tasaki for answering some technical questions on quantum spin systems and help in finding a number of references. 
\newpage
\section{Models}\label{Smodels}
To make our  point, it suffices to treat the \emph{quantum Ising chain} in the thermodynamic limit with (quasi-) local observables, 
the  \emph{quantum Curie--Weisz model} in the same limit with global observables, 
and the \emph{symmetric double well} potential  in the classical limit. For pedagogical reasons we start with the latter, which is much easier to understand; cf.\ Landsman \& Reuvers (2013) for a more detailed treatment from our current perspective. 
\subsection{Double well }
The quantum double-well Hamiltonian on the real axis  is given by
\begin{equation}
H^{DW}_{\hbar}=-\frac{\hbar^2}{2m}\frac{d^2}{dx^2}+\quar \lm (x^2-a^2)^2, \label{TheHam2}
\end{equation}
defined as an unbounded operator on the \Hs
 \begin{equation}
\H^{DW}_{\hbar}=L^2(\R);
\end{equation}
 more precisely, on a domain like $C_c(\R)$ or the Schwartz space of test functions $\CS(\R)$, where it is essentially self-adjoint (Reed \&\ Simon, 1978). We assume  $\lm>0$, and $a>0$.
 
Reflection in the origin of the position coordinate endows this model with a $\Z_2$  symmetry, which is implemented by the unitary operator $u:L^2(\R)\raw L^2(\R)$ defined by
\begin{equation}
 u\Psi(x)=\Psi(-x). \label{uz2}
\end{equation}
The $\Z_2$-symmetry of the Hamiltonian then reads $[H^{DW}_{\hbar},u]=0$, or, equivalently,
\begin{equation}
 uH^{DW}_{\hbar}u^*=H^{DW}_{\hbar}.
\end{equation}

Although this is not necessary for a mathematically correct treatment of this quantum system,  in order to 
better understand the classical limit $\hbar\raw 0$ as well as to see the analogy with our other two models from quantum statistical mechanics, it is convenient to model the double-well system through 
the (admittedly idealized\footnote{Any bounded operator may be obtained as a weak or strong limit of some sequence of compact operators, whereas any possibly unbounded self-adjoint operator resurfaces as a generator of some unitary (hence bounded!)  \rep\ of $\R$ as an additive group, as in Stone's Theorem. This sort of idealization by mathematical convenience is unrelated to the problems involved in the idealizations $\hbar=0$ (i.e., classical physics) or $N=\infty$ (infinite systems) that form the main subject of this paper.}) algebra of observables
\begin{equation}
\CA_{\hbar}=K(L^2(\R)), \: \hbar>0, \label{defAhh}
\end{equation}
i.e., the \ca\ of compact operators on $L^2(\R)$. Algebraically, time-evolution is given by a (strongly continuous) group homomorphism $\ta:\R\raw \mathrm{Aut}(\CA_{\hbar})$ from the time-axis $\R$ (as an additive group) to the group of all  automorphism of $\CA_{\hbar}$,\footnote{
An automorphism of a \ca\ $\CA$ is an invertible linear map $\al:\CA\raw\CA$ satisfying $\al(ab)=\al(a)\al(b)$ and $\al(a^*)=\al(a)^*$. It is \emph{inner} if there is unitary element $u\in\CA$ such that $\al(a)=uau^*$ for all $a\in\CA$.}  written  $t\mapsto \ta_t$; we sometimes abbreviate  $a(t)\equiv \ta_t(a)$. In the model at hand, for any $\hbar>0$ we have
\begin{equation}
\ta^{(\hbar)}_t(a)=u^{(\hbar)}_t a(u^{(\hbar)}_t)^*, \label{ate}
\end{equation}
where the unitary operators $u^{(\hbar)}_t$ are given by
\begin{equation}
u^{(\hbar)}_t=e^{itH^{DW}_{\hbar}/\hbar}. \label{utDW}
\end{equation}

Similarly, the $\Z_2$-symmetry of the model is algebraically described by a group homomorphism $\til{\gm}:\Z_2\raw \mathrm{Aut}(\CA_{\hbar})$, where $\gm(1)=\mathrm{id}$, whilst  the nontrivial element $-1$ of $\Z_2$ is mapped to the automorphism $\gm\equiv \til{\gm}(-1)$ defined in terms of  the unitary $u$ in \er{uz2}
by
\begin{equation}
\gm(a)=uau^*.
\end{equation}
For  $\hbar>0$, the $\Z_2$-invariance of the model is then expressed algebraically by the  property
\begin{equation}
\ta^{(\hbar)}_t\circ \gm=\gm\circ\ta^{(\hbar)}_t,\: t\in\R. \label{circ1}
\end{equation}

The classical limit of this system is a particle  in phase space $\R^2$, with Hamiltonian
\begin{equation}
h^{DW}(p,q)=\frac{p^2}{2m}+\quar \lm (q^2-a^2)^2.\label{Hcl}
\end{equation}
The classical $\Z_2$-symmetry is  given by  the  map  
$\gm_*:(p,q)\mapsto(p,-q)$ on phase space $\R^2$, with pullback $f\mapsto \gm_*^*f\equiv \gm^{(0)}f$ on functions $f$ on $\R^2$, i.e., $\gm^{(0)}f(p,q)=f(p,-q)$. Clearly, 
\beq
 \gm^{(0)}h^{DW}=h^{DW}.
\eeq 
Algebraically,  we take the (idealized) classical  observables to be
\begin{equation}
\CA_0=C_0(\R^2),\label{defAh0}
\end{equation}
i.e., the commutative \ca\ of continuous functions $f:\R^2\raw\C$ that vanish at infinity (with pointwise multiplication). Time-evolution $t\mapsto \ta_t^{(0)}$ on $\CA_0$ is defined by
\beq
\ta_t^{(0)}f(p,q)=f(p(t),q(t)),
\eeq 
where $(p(t),q(t))$ is the solution of the Hamiltonian equations of motion following from \er{Hcl}, with initial conditions $(p(0),q(0))=(p,q)$. Then $\Z_2$-invariance is expressed by 
\begin{equation}
\ta_t^{(0)}\circ  \gm^{(0)}= \gm^{(0)}\circ \ta_t^{(0)},\: t\in\R.\label{circ0}
\end{equation}
\subsection{Quantum Ising chain}\label{QICSec}
For $N<\infty$, the  Hamiltonian of the quantum Ising chain (with $J=1$ for simplicity) is
\begin{equation}H^I_N=-\sum_{i=-\half N}^{\half N-1}\sg^z_i \sg^z_{i+1}-B\sum_{i=1}^N \sg^x_i, \label{qising}
\end{equation}
where $N$ is  even and  and we adopt  free boundary conditions. This operator is  defined on 
\begin{equation}
\H_N=\ot^N\C^2,\label{otH}
\end{equation}
i.e., the $N$-fold tensor product of $\C^2$, where we label the first copy with $j=-\half N$, the second with $j=-\half N+1$, \ldots, and the last one with $j=\half N-1$.\footnote{
The Pauli matrix  
$\sg_i^{\mu}\equiv \otimes_{j=-\half N}^{i-1}1_2\otimes\sg^{\mu}\otimes_{k=i+1}^{\half N-1} 1_2\:\: (i=1,\ldots, N,\, \mu=x,y,z)$ 
 acts on the $i$'th  $\C^2$.}
 We assume $B>0$.
   This model describes a chain of $N$ immobile spin-$\half$ particles with ferromagnetic coupling in a transverse magnetic field (Pfeuty, 1970; Sachdev, 2011; Suzuki et al, 2013).\footnote{The quantum Ising model is a special case of the $XY$-model, to which the same conclusions apply.} 
 
 Although the physically relevant operators (like the above Hamiltonian) are most simply given if the \Hs\ is realized in the tensor product form \er{otH}, the physical interpretation of states tends to be more transparent if we realize $\H_N$ as $\ell^2(S_N)$,\footnote{For any countable set $S$, the \Hs\ $\ell^2(S)$ consists of all  functions $f:S\raw\C$ that satisfy $\sum_{s\in S} |f(s)|^2<\infty$, with inner product $\la f,g\ra=\sum_{s\in S} \ovl{f(s)}g(s)$. Of course, for $N<\infty$, $S_N$ is a finite set.}
  where $S_N=\underline{2}^{\underline{N}}$ is the space of classical spin configurations $s:\underline{N}\raw\underline{2}$. Here 
  \begin{eqnarray}
       \underline{2}&=&\{-1,1\};\\
  \underline{N}&=&\{-\half N,\half N-1\}\:\: (N\geq 2).
\end{eqnarray}
 In terms of the standard basis $|1\ra=(1,0)$ and $|-1\ra=(0,1)$ of $\C^2$, where the label $|\lm\ra$ is the corresponding eigenvalue of $\sg^z$, a suitable unitary equivalence $v_N:\ell^2(S_N)\raw \ot^N\C^2$ (each isomorphic to $\C^{2^N}$) is given by linear extension of
\begin{equation}
v_N \dl_s=  \otimes_{j=-\half N}^{\half N-1}|s(j)\ra,\:\: s,t\in S_N,
\end{equation}
where $\dl_s$ is defined by $\dl_s(t)=\dl_{st}$; such functions form an orthonormal basis of  $\ell^2(S_N)$. 

For example, the state with all spins up, i.e.,
$\ot^N |1\ra$, corresponds to $\dl_{s_{\uaw}}$, where $s_{\uaw}(j)=1$ for all $j$, and analogously  $s_{\dow}(j)=-1$ for all $j$ is the state with all spins down.
Thus the advantage of this realization is that we may talk of localization of states in spin configuration space, in the sense that some $\Psi\in \ell^2(S_N)$ may be peaked on just a few spins configurations (whilst typically having small but nonzero values elsewhere).

For any  $B$,  the quantum Ising chain has a $\Z_2$-symmetry given by a 180-degree rotation around the $x$-axis. This symmetry is implemented by the unitary operator 
\begin{equation}
u_N=\ot_{i\in\underline{N}} \sg_i^x
\end{equation}
on $\H_N$, which satisfies $[H^I_N,u_N]=0$, or, equivalently,\footnote{Note that
 $u_N\sg_i^xu_N^*=  \sg_i^x$, $u_N\sg_i^yu_N^*= -\sg_i^y$, $u_N\sg_i^zu_N^*= -\sg_i^z$, which implies \er{2.11}.}
\beq
u_NH^I_Nu_N^*= H^I_N.
\label{2.11}
\eeq

 As in the previous subsection, we could take a purely algebraic approach by defining the \ca\  of observables of the system for $N<\infty$ to be
\begin{equation}
\CB_N=\ot^N M_2(\C), \label{defAN}
\end{equation}
i.e., the $N$-fold tensor product  of the $2\x 2$ matrices, labeled as described after \er{otH}.
 Time-evolution is  given by the analogue of \er{ate}, which explicitly reads
 \begin{eqnarray}
\ta^{(N)}_t(a)&=&u_t^{(N)}a(u_t^{(N)})^*;\label{usual1}\\
u_t^{(N)}&=&e^{itH^I_N}.\label{usual2}
\end{eqnarray}
Furthermore, the $\Z_2$-symmetry is given by the automorphism $\gm_N$ of $\CB_N$ defined by
\beq\gm_N(a)= u_Nau_N^*, \: a\in \CB_N.\eeq 
In this language, $\Z_2$-invariance of the model is expressed as in \er{circ1}, viz.
\begin{equation}
\ta_t^{(N)}\circ \gm_N=\gm_N\circ\ta^{(N)}_t,\: t\in\R. \label{circ2}
\end{equation}

Without taking the dynamics into account, given \er{defAN} there are two (interesting) possibilities for the limit algebra at $N=\infty$ (Landsman, 2007, \S 6):
\begin{itemize}
\item The C*-algebra of (quasi-)\emph{local} observables is 
\beq
\CB^l_{\infty}=\ot^{\Z} M_2(\C), \label{Aloc}
\eeq  the infinite tensor product of $M_2(\C)$ as defined in Kadison \& Ringrose (1986, \S11.4);
\item The C*-algebra of \emph{global} observables is 
\beq \CB^g_{\infty}=C(\CS(M_2(\C))), \label{Aglob}
\eeq
the  C*-algebra of continuous functions on the state space $\CS(M_2(\C))$
of $M_2(\C)$.
\end{itemize}
Thus $\CB^l_{\infty}$  is highly noncommutative, like each of the $\CB_N$, which is embedded in
$\CB^l_{\infty}$ by tensoring  with infinitely many unit matrices in the obvious way, whereas 
 $\CB^g_{\infty}$ is obviously commutative (under pointwise multiplication, that is). 
Which of these limit algebras is the appropriate one depends on the Hamiltonian: for short-range interactions, as in \er{qising}, it is the  first, because the finite-$N$ Hamiltonians induce a well-defined time-evolution on  $\CB^l_{\infty}$.  For mean-field models
like the quantum Curie--Weisz model, on the other hand, it is the second, for the same reason; see below and the next subsection, respectively.

Indeed, in the first case, for Hamiltonians like
\er{qising}, by Theorem 6.2.4 in   Bratteli \&\  Robinson (1997) there exists a unique  time-evolution 
$\ta$ on $\CB^l_{\infty}$, again in the sense of a strongly continuous group homomorphism $\ta:\R\raw \mathrm{Aut}(\CB^l_{\infty})$, that
extends the local dynamics given by the local Hamiltonians $H^I_N$ in that 
\begin{equation}
\ta_t(a)=\lim_{N\raw\infty}\ta^{(N)}_t(a), \: a\in \CB^l_{\infty}. \label{brresult}
\end{equation}
Equivalently,  for each  $a\in \CB^l_{\infty}$ that is \emph{local} in being contained in some $\CB_N\subset \CB^l_{\infty}$,
  \begin{equation}
\frac{da(t)}{dt}=i\lim_{N\raw\infty} [H^I_N,a(t)],\label{delta0}
\end{equation}
where $a(t)\equiv \ta_t(a)$.
 So the limit theory $\mathsf{H}$  is the pair $(\CB^l_{\infty},\ta)$, in which
the local Hamiltonians $H_N^I$ have been replaced by the single one-parameter automorphism group $\ta$.

Finally, the infinite-volume relic of the $\Z_2$-symmetry $u_N$ is the automorphism $\gm$ of $\CB^l_{\infty}$ that is uniquely defined by the property  $\gm(a)=u_Nau_N^*$
for each $a\in \CB_N$, cf.\ \er{2.11}.
The invariance property \er{2.11} of the local Hamiltonians
then becomes the $\Z_2$-symmetry of the time-evolution $\ta$, as expressed by \er{circ1} (\emph{mutatis mutandis}).
\subsection{Quantum Curie-Weisz model}
For $N<\infty$, the Hamiltonian of the quantum Curie-Weisz model is
\begin{equation}
H^{CW}_N=-\frac{1}{2N}\sum_{i,j=-\half N}^{\half N-1}\sg^z_i \sg^z_{j}-B\sum_{i=1}^N \sg^x_i, \label{CW}
\end{equation}
acting on the same \Hs\ $\H_N$ as the  Hamiltonian \er{qising}, i.e., \er{otH}, but differing from it by the spin-spin interaction  being nonlocal and even of arbitrary range.\footnote{
The name \emph{Lipkin model} may be found in the nuclear physics literature, and without the $B$-term, \er{CW} is sometimes called the \emph{Weisz model}.
It may be treated in any spatial dimension in much the same way. In the context of the measurement problem, the quantum Curie-Weisz model has been extensively studied by Allahverdyana,  Balian, \&\  Nieuwenhuizen (2013), with hardly any overlap with our analysis though.} 

In terms of the marcoscopically averaged spin operators
\begin{equation}
S_{\mu}=\frac{1}{2N}\sum_{i=-\half N}^{\half N-1}\sg^{\mu}_i, \: \mu=1,2,3, \label{maspin}
\end{equation}
this Hamiltonian assumes the perhaps more transparent form
\begin{equation}
H^{CW}_N=-2N(S_z^2+BS_x). \label{CW2}
\end{equation}
This model has exactly the same $\Z_2$-symmetry as its local counterpart \er{qising}, and for finite $N$ one can simply copy all relevant formulae from the previous subsection.

What is markedly different for long-range forces (as opposed to local ones), however, is the correct choice of the limit algebra, which (in the context of similar models, starting from the {\sc bcs} Hamiltonian for superconductivity) emerged from the work of
Bogoliubov, Haag, Thirring, Bona, Duffield, Raggio, Werner, and others; see Landsman (2007, Ch.\ 6) for references, as well as for the reformulation in terms of continuous fields of \ca s to be given in  \S\ref{cf} below. To make a long story short, the local Hamiltonians \er{CW2} do not induce a time-evolution on $\CB^l_{\infty}$, but they do so on the commutative algebra \er{Aglob}. 

This limiting dynamics turns out to be of Hamiltonian form, as was to be expected for a classical theory, albeit
of a generalized form, where the underlying phase space is a Poisson manifold that is not symplectic, see e.g. Landsman (1998). Specifically, we use the fact that  the state space $\CS(M_2(\C))$
of $M_2(\C)$ appearing in \er{Aglob} is isomorphic (as a compact convex set) to the 
 three-ball $\mathsf{B}^3$, which consists of all $(x,y,z)\in\R^3$ satisfying $x^2+y^2+z^2\leq 1$. The isomorphism in question is given by the well-known parametrization \begin{equation}
\rh(x,y,z)=\half   \left(\begin{array}{cc} 1+z & x-iy \\ x+iy & 1-z \end{array}\right),
\label{gens2} \end{equation}
of an arbitrary density matrix on $\C^2$. This parametrization is such that pure states (i.e., those for which $\rh$ is a one-dimensional projection, so that $\rh^2=\rh$) are mapped to the boundary $S^2$ of $\mathsf{B}^3$, whose points satisfy $x^2+y^2+z^2\leq 1$ (this is just the familiar Bloch sphere from physics texts). Now $\R^3$ is equipped with (twice) the so-called Lie--Poisson bracket,\footnote{See e.g.\ Marsden \& Ratiu (1994); our phase space $\R^3$ is the dual of the Lie algebra of $SO(3)$. The factor 2 in \er{tlp} is caused by the fact that 
our model has basic spin one-half (or use $x_i'=\half x_i$ to avoid it).}
which on the coordinate functions $(x_1,x_2,x_3)\equiv (x,y,z)$  is  given by
\begin{equation}
\{x_i,x_j\}=-2\varep_{ijk}x_k, \label{tlp}
\end{equation}
and is completely defined by this special case through the Leibniz rule. In terms of this bracket, time-evolution is given by the familiar Hamiltonian formula 
\begin{equation}
\frac{dx_i}{dt}=\{h^{CW},x_i\}, \label{fam}
\end{equation}
where the Hamiltonian giving the limiting dynamics of \er{CW2} is given by
\begin{equation}
h^{CW}(x,y,z)=-(\half z^2+Bx). \label{hcwc}
\end{equation}
Thus the equations of motion \er{fam} are given by
\begin{equation}
\frac{dx}{dt}=2yz;\:\:\:\frac{dy}{dt}=2z(B-x); \:\:\: \frac{dz}{dt}=-2By.
\end{equation}
\section{Continuity}\label{cf}
It is often stated that limits like $N\raw\infty$ and  $\hbar\raw 0$ are `singular', and indeed they are, if one merely looks at the Hamiltonian: putting $\hbar=0$ in \er{TheHam2} yields an operator that has practically nothing to do with the small $\hbar$-behaviour of its originator, letting $\hbar\raw 0$ in \er{utDW} seems to make no sense, and the limit $N\raw\infty$ of \er{qising} or \er{CW} is simply undefined. Nonetheless, \emph{the limits in question are continuous} if treated in the right way.\footnote{The following discussion amplifies our earlier treatments of the  $\hbar\raw 0$, $N\raw\infty$ limits in Landsman (1998, 2007), respectively, by including dynamics and, in the next few sections, ground states and \ssb.} 

They key, which may be  unexpected at first sight, is that in discussing the $\hbar\raw 0$ limit of \qm\ it turns out to be possible to ``glue'' the (highly noncommutative!) algebras of quantum observables
$\CA_{\hbar}$ in \er{defAhh} continuously to the (commutative!) algebra of classical observables $\CA_0$ in \er{defAh0}, whereas for the $N\raw\infty$ limit of quantum statistical mechanics there are even two (relevant) possibilities: the (noncommutative) algebras $\CB_N$ in \er{defAN} can be glued continuously to either 
the algebra of (quasi-) local observables $\CB^l_{\infty}$ in \er{Aloc}, which is 
noncommutative, too (and hence is the more obvious choice), or to the commutative algebra of global observables $\CB^g_{\infty}$ in \er{Aglob}. The ``glueing'' is done using the formalism of \emph{continuous fields of \ca s} (of observables); we  will just look at the cases of interest for our three models, and refer the reader to Dixmier (1977) and  Kirchberg \& Wassermann (1995) for the abstract theory.\footnote{The idea of a continuous fields of \ca s goes back to Dixmier (1977), who gave a direct definition in terms of glueing conditions between the fibers, and was usefully reformulated by Kirchberg \& Wassermann (1995), who stressed the role of the continuous sections of the field already in its definition. Both definitions are reviewed in Landsman (1998). Our  informal discussion below uses elements of both.} 
Continuity of the dynamics in the various limits at hand will be a corollary, provided time-evolution is expressed in terms of the one-parameter automorphism groups $\ta$. Finally, the ensuing  \emph{continuous fields of states} defined by the continuous fields of observables will provide the right language for the technical elaboration of our solution of the problem of emergence, as described in the Introduction (N.B.: the use of this formalism itself is \emph{not} yet the solution!).
\subsection{Continuous fields of \ca s}\label{cfcas}
Each of our three continuous fields $\mathfrak{A}^c$, $\mathfrak{A}^l$, and $\mathfrak{A}^g$ of interest
is defined over  a (`base') space $I_{\al}\subseteq [0,1]$ defined  for $\al=c,l,g$
as  $I_c=[0,1]$ and $I_l=I_g=\{0\}\cup1/(2\N_*)$,\footnote{That is, $x\in I_l$ if either $x=0$ or $x=1/N$ for some even $N\in\N\backslash\{0\}$.} 
 with the topology inherited from $[0,1]$ (in which each $I_{\al}$ is compact). The fibers are  as follows:
\begin{enumerate}
\item   $\mathfrak{A}^c_{\hbar}=\CA_{\hbar}=K(L^2(\R))$ for  $h\in (0,1]$ and $\mathfrak{A}^c_0=\CA_0=C_0(\R^{2})$;
\item  $\mathfrak{A}^l_{1/N}=\CB_N=\ot^N M_2(\C)$  for even $N<\infty$ and $\mathfrak{A}^l_0=\CB^l_{\infty}=\ot^{\Z} M_2(\C)$;
\item  $\mathfrak{A}^g_{1/N}=\CB_N=\ot^N M_2(\C)$  for even $N<\infty$ and $\mathfrak{A}^g_0=\CB^g_{\infty}=C(\CS(M_2(\C)))$.\end{enumerate}
The continuity structure is given by specifying a sufficiently large family of continuous cross-sections, i.e., maps $\sg$ assigning some element $a\in \mathfrak{A}^{\al}_x$ to each $x\in I_{\al}$, as follows.\footnote{Lest this operation may appear to be circular: there are  stringent and highly exclusive conditions on admissible continuity structures, namely, for  continuous cross-sections $\sg$ of a continuous field $\mathfrak{A}^{\al}$ over  $I_{\al}$:
\begin{enumerate}
\item
The function $x\mapsto  \|\sg(x)\|_{x}$, where $\|\cdot\|_{x}$ is the norm in $\mathfrak{A}_x$, is continuous (from  $I_{\al}$ to $\R$);
\item
For any $f\in C(I_{\al})$ and $\sg$ as above the  cross-section
$x\mapsto f(x)\sg(x)$  for all $x\in I_{\al}$ is again continuous;
\item The totality of all $\sg$ form a \ca\ in the norm $\|\sg\|=\sup_x \|\sg(x)\|_x$ and pointwise operations.
\end{enumerate}
}
\begin{enumerate}
\item Each $f\in C_0(\R^{2}$ defines an operator $Q_{\hbar}(f)\in K(L^2(\R))$ by \emph{Berezin quantization},
\beq
 Q_{\hbar}(f)=\int_{\R^{2}} \frac{dp
dq}{2\pi\hbar}\, f(p,q) | \Phi^{(p,q)}_{\hbar}\rangle\langle\Phi^{(p,q)}_{\hbar}|,
\eeq
where, for any unit vector $\Psi$, the one-dimensional projection onto $\C\cdot\Psi$ is denoted by $|\Psi\ra\la\Psi|$, and the  coherent states $\Phi^{(p,q)}_{\hbar}\in L^2(\R)$, $(p,q)\in\R^2$, are defined by 
\beq
\Phi^{(p,q)}_{\hbar}(x)=(\pi\hbar)^{-1/4}e^{-
ipq/2\hbar}e^{ipx/\hbar}e^{-(x-q)^2/2\hbar}.\label{pqcohst} 
\eeq
Using this quantization map, for each $f$ we then define a cross-section $\sg_f$ of $\mathfrak{A}^c$ by 
\begin{eqnarray}
\sg_f(0)&=&f;\\
 \sg_f(\hbar)&=&Q_{\hbar}(f), \: \hbar\in (0,1].
\end{eqnarray}
Thus, although $f$ and $Q_{\hbar}(f)$ are completely different mathematical objects, for small $\hbar$ they are 
 sufficiently close to each other to be able to say that $\lim_{\hbar\raw 0} Q_{\hbar}(f)=f$, in the sense that if one continuously follows the curve $\hbar\mapsto \sg_f(\hbar)$ in the total space $\coprod_{\hbar\in[0,1]} \mathfrak{A}^c_{\hbar}$ of the bundle (equipped with the topology that makes this disjoint union a continuous field of \ca s) all the way down to $\hbar=0$, one ends up with $f$.
  \item In order to describe  this case, we have to realize the infinite tensor product $\CB^l_{\infty}$ as equivalence classes of quasi-local sequences (Raggio \& Werner, 1989):
  \begin{enumerate}
\item   A sequence $(a)\equiv (a_N)_{N\in 2\N_*}$ is \emph{local} if there is an $M$ such that $a_N=\iota_{MN}(a_M)$ for all $N\geq M$, where $\iota_{MN}:\CB_M\hookrightarrow\CB_N$ is the inclusion map (which takes the tensor product of $a_M\in\CB_M$ with as many unit matrices as needed to make it an element of $\CB_N$).
\item A sequence $(a)$ is \emph{quasi-local} if for any $\varep>0$ there is an $M$ and a local sequence $(a')$ such that $\|a_N-a'_N\|<\varep$ for all $N>M$ (cf.\ \S\ref{twosec} around \er{limitf}).
\item Introduce an equivalence relation on the quasi-local sequences by saying that $(a)\sim(a')$ if  $\lim_{N\raw\infty}\|a_N-a'_N\|=0$. 
\item $\CB^l_{\infty}$ consists of equivalence classes $[a]\equiv a_{\infty}$ of quasi-local sequences; these form a \ca\ under pointwise operations (in $N$) and norm
  $\|a_{\infty}\|=\lim_{N\raw\infty}\|a_N\|$.
\end{enumerate}
Continuous cross-section of  $\mathfrak{A}^l$ then correspond to quasi-local sequences $(a)$ through
\begin{eqnarray}
 \sg(1/N)&=&a_N;\label{csg1}\\
 \sg(0)&=&a_{\infty}. \label{csg2}
\end{eqnarray}
  \item Here, the embedding maps $\iota_{MN}$ are replaced by the obvious symmetrization maps 
  $j_{MN}:\CB_M\hookrightarrow\CB_N$, defined for $N>M$ by $j_{MN}=\mathrm{Sym}_N\circ \iota_{MN}$, where the usual symmetrization map $\mathrm{Sym}_N:\CB_N\raw\CB_N$ projects on the completely symmetric tensors. 
  \begin{enumerate}
\item  A sequence $(a)$ is \emph{symmetric}  if there is an $M$ such that $a_N=j_{MN}(a_M)$ for all $N\geq M$.
\item A sequence $(a)$ is  \emph{quasi-symmetric}  if for any $\varep>0$ there is an $M$ and a symmetric sequence $(a')$ such that $\|a_N-a'_N\|<\varep$ for all $N>M$.
\end{enumerate}
 If $(a)$ is a quasi-symmetric sequence, and $\om$ is a state on $M_2(\C)$, then the following limit exists:
 \begin{equation}
a_0(\om)=\lim_{N\raw\infty} \om^N(a_N), \label{defa0}
\end{equation}
 where $\om^N$ is the $N$-fold tensor product of $\om$ with itself, defining a state on $\CB_N$. The ensuing function $a_0$ on the state space $\CS(M_2(\C))$ is  continuous, so that $a_0$ is an element of the limit algebra $\CB^g_{\infty}$. If we identify $\CS(M_2(\C))$ with $\mathsf{B}^3$, then from \er{gens2},
 \begin{equation}
a_0(x,y,z)=\lim_{N\raw\infty}\Tr\left(\rh(x,y,x)^{\otimes N} a_N\right). \label{coma0}
\end{equation}
The continuous cross-sections of  $\mathfrak{A}^g$ are then simply given by maps $\sg$ of the form
\begin{eqnarray}
 \sg(1/N)&=&a_N;\label{csl1}\\
 \sg(0)&=&a_0,\label{csl2}
\end{eqnarray}
the latter defined as in \er{defa0}, where $(a)$ is some quasi-symmetric sequence.
  \end{enumerate}
  One may be surprised by the commutativity of $\mathfrak{A}^g_0$, but a simple example may clarify this (cf.\ Landsman (2007, \S6.1) for the general argument). If we write the left-hand side of \er{maspin} as $S_{\mu}^{(N)}$ for clarity, then $a_N=S_{\mu}^{(N)}=\half j_{1N}(\sg^{\mu})$ defines a symmetric sequence, and 
   \beq
   [S^{(N)}_{\mu}, S^{(N)}_{\nu}]=\frac{i}{N}\varep_{\mu\nu\rh}S^{(N)}_{\rh}.
   \eeq 
   This commutator evidently vanishes as $N\raw\infty$. The limit functions $a_0$ may be computed from 
\er{coma0}. Writing $\til{S}_{\mu}$ for the limit of the sequence $a_N=S_{\mu}^{(N)}$, we obtain
\begin{equation}
\til{S}_{\mu}(x_1,x_2,x_3)=\half x_{\mu}.
\end{equation}
Similarly, the mean (i.e., averaged) Hamiltonians 
\begin{equation}
h^{CW}_N=H^{CW}_N/N=-2(S_z^2+BS_x)
\end{equation}
 of the Curie-Weisz model,\footnote{And similarly for general mean-field models, see Bona (1988) and Duffield \& Werner (1992).} cf.\ \er{CW2}, define a symmetric sequence, with limit \er{hcwc}, i.e., 
\begin{equation}
\til{h}^{CW}=h^{CW}.
\end{equation}
  \subsection{Continuous fields of states}\label{Cfs}
  Dually, one has \emph{continuous fields of states}, which, given one of our continuous fields of \ca s 
  $\mathfrak{A}^{\al}$,  are simply defined as families $(\om)=(\om_x)_{x\in I_{\al}}$, where $\om_x$ is a state on 
  $\mathfrak{A}^{\al}_x$, such that for each continuous cross-section $\sg$ of   $\mathfrak{A}^{\al}$, the function
 $x\mapsto \om_x(\sg(x))$ is continuous on $I_{\al}$. The main purpose of this is to define limits (cf.\ \S\ref{GS}). In our examples:
 \begin{enumerate}
\item For the classical limit this reproduces the standard notion of convergence of quantum states to classical ones,
which is as follows.\footnote{Cf.\ Robert (1987), Paul \& Uribe (1996), Landsman \& Reuvers (2013), and many other sources.}
Let $(\rh_{\hbar})$ be a family of  density matrices  on $L^2(\R)$, $\hbar\in(0,1]$. Each $\rh_{\hbar}$ defines
a probability measure $\mu_{\hbar}$ on phase space $\R^2$ by
\begin{equation}
\int_{\R^2} d\mu_{\hbar}\, f= \Tr (\rh_{\hbar}  Q_{\hbar}(f)). \label{defmuh}
\end{equation}
A state $\rh_0$ on $C_0(\R^2)$ equally well defines an associated
 measure $\mu_0$ by Riesz--Markov,
 \beq\rh_0(f)=\int_{\R^2} d\mu_0\, f.\label{RMT}\eeq
The family $(\rh_{\hbar})$ then converges to $\rh_0$ iff $\mu_{\hbar}\raw\mu_0$ weakly, in that for each $f\in C_c(\R^2)$, 
\beq
\lim_{\hbar\raw 0} \int_{\R^2} d\mu_{\hbar}\, f=\int_{\R^2} d\mu_0\, f.\eeq 
 \item Any state $\om_0$ on $\mathfrak{A}^l_0=\CB^l_{\infty}$ defines  a state
$\om_{1/N}$ on $\mathfrak{A}^l_{1/N}=\CB_N$ by restriction (since $\CB_N\subset \CB^l_{\infty}$), and the ensuing field of states $(\om)$ on 
 $\mathfrak{A}^l$ is (tautologically) continuous.
 
  Conversely, any continuous field of states $(\om')$ on  $\mathfrak{A}^l$ is asymptotically equal to one of the above kind, in that the field $(\om)$ defined by $\om_0=\om'_0$ has the property that $\lim_{N\raw\infty} |\om_{1/N}(a_N)- \om'_{1/N}(a_N)|=0$ for any fixed quasi-local sequence $(a)$.\footnote{Indeed, for any $\varepsilon>0$, there is an $M$ such that   $|\om_{1/N}(a_N)- \om'_{1/N}(a_N)|<\varepsilon$ for all $N>M$.}
  \item For $\mathfrak{A}^g$, no independent characterization of continuous fields of states seems available.
  A nice example, though, comes from permutation-invariant states $\om^l$ on $\CB_{\infty}^l$ (no typo), defined by the property that 
   each restriction $\om^l_{1/N}=\om^l_{|\CB_N}$ to $\CB_N$ is permutation-invariant. These $\om^l_{1/N}$ define a  continuous field of states on $\mathfrak{A}^g$, whose limit state $\om_0^l$ on $\mathfrak{A}^g_0=C(\mathsf{B}^3)$ yields a probability measure $\mu_0^l$, which even characterizes the original state  by  the quantum De Finetti Theorem of  St\o rmer (1969):
    \beq
    \om^l=\int_{\mathsf{B}^3} d\mu_0^l(x,y,z)\, \rh(x,y,z)^{\infty}.
    \eeq Here we identify the density matrix \er{gens2} with the corresponding state on $M_2(\C)$ via the trace pairing, see \er{coma0}, and $\rh^{\infty}=\lim_{N\raw\infty}\rh^N$, as in \er{defa0}.
    
Of course, any  continuous fields of states $(\om)$ on $\mathfrak{A}^g$ defines  a probability measure $\mu_0$ on $\mathsf{B}^3$ by applying the Riesz--Markov Theorem to the limit state
 $\om_0$. We will see various interesting examples of such measures in the remainder of this paper.\footnote{See also  Landsman (2007, \S6.2) for a number of abstract general results on such limit measures.} 
\end{enumerate}
 \subsection{Continuity of time-evolution}
 Using the continuous field picture developed above, we also obtain a satisfactory notion of convergence of time-evolution in our various limits. We list our three cases of interest.\footnote{Vast generalizations of the material are possible, but we restrict attention to our three guiding models.} For technical reasons the optimal result, where time-evolved continuous cross-section of $\mathfrak{A}^{\al}$ are again continuous, applies only to $\al=l,g$,
 but for $\al=c$ one still has a weaker continuity result after pairing  continuous cross-sections with continuous fields of states. 
\begin{enumerate}
\item Let $(\rh_{\hbar})_{\hbar\in[0,1]}$ be a continuous family of states on $\mathfrak{A}^c$, where for $\hbar>0$ we identify the state with the associated density matrix,
 with associated probability measures  $\mu_{\hbar}$, defined by \er{defmuh}, so that $\lim_{\hbar\raw 0}\mu_{\hbar}=\mu_0$ weakly. Each density matrix $\rh_{\hbar}$ evolves in time according to the Liouville--von Neumann equation determined by the quantum Hamiltonian \er{TheHam2}. In other words, $\rh_{\hbar}(t)$ satisfies
\begin{equation}
\Tr (\rh_{\hbar}(t)a)=\Tr (\rh_{\hbar}\ta_t^{(\hbar)}(a)), \: a\in K(L^2(\R)).
\end{equation}
This induces a time-evolution on each $\mu_{\hbar}$, in that $\mu_{\hbar}(t)$ is the probability measure determined by $\rh_{\hbar}(t)$ according to \er{defmuh}.

Likewise,
$\mu_0$  evolves in time according to the classical Liouville equation given by the classical Hamiltonian
\er{Hcl}, which is measure-preserving and hence also maps $\mu_0$ into some other probability measure $\mu_0(t)$.
For unbounded Hamiltonians of the kind \er{TheHam2}, the best continuity result is then given by the weak limit\footnote{This  follows from Egorov's Theorem in the form of Theorem II.2.7.2 in Landsman (1998). In the `academic' case that the quantum Hamiltonian  is compact and is given by $H_{\hbar}=Q_{\hbar}(h)$,
one has operator convergence of the kind $\lim_{\hbar\raw 0} \|Q_{\hbar}(\ta_t^{(0)}(f))-\ta_t^{(\hbar)}(Q_{\hbar}(f))\|=0$, see Prop.\ 2.7.1 in Landsman (1998). This also implies that time-evolved continuous cross-section of $\mathfrak{A}^c$ are again continuous, which unfortunately does not seem to be the case for the unbounded Hamiltonian  \er{TheHam2}. }
\begin{equation}
\lim_{\hbar\raw 0} (\mu_{\hbar}(t))=\mu_0(t), \: t\in\R. \label{pointt}
\end{equation}
\item The quasi-local case leads to  stronger results, since the operator limit \er{brresult} implies
\begin{equation}
\ta_t(a_{\infty})=\lim_{N\raw\infty}\ta^{(N)}_t(a_N), \: t\in\R, \label{brresult2}
\end{equation}
for any quasi-local sequence $(a)$.\footnote{This holds within $\CB^l_{\infty}$ in the operator norm, where
each $\ta^{(N)}_t(a_N)$ is embedded in $\CB^l_{\infty}$.}
 In other words, if $(a)$ is a quasi-local sequence with limit $a_{\infty}$ , then the
 time-evolved quasi-local sequence $(a(t))$, where each $a_N$ evolves with $\ta^{(N)}_t$,
  is again quasi-local, with limit $a_{\infty}(t)$, where $a_{\infty}$ evolves with the time-evolution $\ta_t$
  directly defined in infinite volume. 
 Equivalently, 
 time-evolved  continuous cross-section of $\mathfrak{A}^l$ are again continuous, where time-evolution is defined separately in each fiber. 
 For continuous fields of states $(\om)$ this implies
\begin{equation}
\lim_{N\raw\infty}(\om_{1/N}(t)(a_N))=\om_0(t)(a_{\infty}), \label{Alcf}
\end{equation}
for any quasi-local sequence $(a)$ with limit $a_{\infty}$, where $\om_{1/N}(t)$ and $\om_0(t)$ are defined by ``Schr\"{o}dinger-picturing'' (that is, dualizing)
$\ta_t^{(N)}$ and $\ta_t$, respectively; cf.\ \er{pointt}.
\item As in the previous case, time-evolved continuous cross-section of $\mathfrak{A}^g$ are again continuous,
but because of the vast difference between the fibers $\mathfrak{A}^g_x$ at $x=1/N$ and at $x=0$, this result may be more unexpected. 
Let $(a)$ be a quasi-symmetric sequence with limit $a_0$,
let $a_N(t)=\ta_t^{(N)}(a_N)$ be defined by the quantum Hamiltonian \er{CW2} in the usual unitary way (i.e., by \er{usual1} and \er{usual2} with $H_N^{CW}$ instead of $H_N^I$), and finally let $a_0(t)$ be defined by the classical Hamiltonian \er{hcwc} and the Poisson bracket \er{tlp}.\footnote{That is, if $\vec{x}(t)$ is the solution of \er{fam} with initial condition $\vec{x}(0)=\vec{x}$, then $a_0(\vec{x})(t)=a_0(\vec{x}(t))$.}
 Then the time-evolved sequence $(a(t))$ is again  quasi-symmetric, with limit $a_0(t)$; see Landsman (2007, \S6.5), elaborating on Duffield \& Werner (1992). Of course, for continuous fields of states this implies a result analogous to \er{Alcf}.
\end{enumerate}
In conclusion, as soon as it has been  (re)formulated in the `right' setting, time-evolution as such turns out to be continuous  throughout the limits $N\raw \infty$ or $\hbar\raw0$.  The precise kind of continuity is even quite strong in our two lattice models, but it is still acceptable for the double well.\footnote{Note that because our three examples display exactly the same anomalies in their ground states (cf.\ the Introduction as well as the next section), the technical differences between these kinds of continuity seem irrelevant to our analysis of the emergence problem.} Given this continuity, the extreme discontinuity in the behaviour of the ground states if one passes from $N<\infty$ to $N=\infty$, or from $\hbar>0$ to $\hbar=0$, is remarkable. Displaying this discontinuity in naked form and at the appropriate technical level is the purpose of the next section; resolving it will be the goal of the one after. 
\section{Ground states}\label{GS}
The definition of a ground state of a specific physical system depends on the setting (i.e., classical/quantum, finite/infinite), but is uncontroversial and well understood in all cases.
\begin{enumerate}
\item A ground state (in the usual sense) of a quantum Hamiltonian like \er{TheHam2}  is a unit eigenvector
 $\Psi_{\hbar}^{(0)}\in L^2(\R)$ of $H_{\hbar}^{DW}$ for which the corresponding eigenvalue $E_{\hbar}^{(0)}$ lies  at the bottom of the spectrum $\sg(H_{\hbar}^{DW})$. Algebraically, such a unit vector $\Psi_{\hbar}^{(0)}$ defines a state $\psi_{\hbar}^{(0)}$ on the \ca\ of observables $\CA_{\hbar}$ given by \er{defAhh}, viz.\footnote{A state on a \ca\ $\CA$ is a positive linear functional  $\om:\CA\raw\C$  of norm one, where positivity of  $\om$ means that $\om(a^*a)\geq 0$ for each $a\in \CA$. If $\CA$ has a unit $1_A$, then $\|\om\|=1$ iff $\om(1_A)=1$ (given positivity).  We say that $\om$ is \emph{pure}  if $\om=p\om_1+(1-p)\om_2$ for some $p\in (0,1)$ and certain states $\om_1$ and $\om_2$ implies $\om_1=\om_2=\om$, and \emph{mixed} otherwise.
  If $\CA=C_0(X)$ is commutative, then by the Riesz--Markov Theorem 
 states on $\CA$ bijectively correspond to probability measures on $X$, the pure states being the Dirac (point) measures $\dl_x$ defined by $\dl_x(f)=f(x)$, $x\in X$. For $\CA=K(L^2(\R))$, the pure states are just the unit vectors.}
 \begin{equation}
\psi_{\hbar}^{(0)}(a)=\la \Psi_{\hbar}^{(0)},a\Psi_{\hbar}^{(0)}\ra, \:\: a\in K(L^2(\R)),
\end{equation}
where $\la\cdot,\cdot\ra$ is the inner product in $L^2(\R)$. This reformulation is useful already for fixed $\hbar$, where it removes the phase ambiguity in unit vectors, but it is mandatory if we wish to take the limit $\hbar\raw 0$.
The above definition of a ground state may be reformulated directly in terms of the algebraic state $\psi_{\hbar}^{(0)}$, see point 2 below. 

 A ground state of the corresponding classical Hamiltonian \er{Hcl} is just a point $z_0\in\R^2$ in phase space where $h^{DW}$ takes an absolute minimum. 
 \item For $N<\infty$, an analogous discussion applies to the quantum Ising Hamiltonian \er{qising}:
 a ground state is simply a unit eigenvector
 $\Psi_N^{(0)}\in \H_N$ of $H_N^I$ whose eigenvalue $E_N^{(0)}$ is minimal, with the appropriate algebraic reformulation (luxurious for $N<\infty$ but needed as $N\raw\infty$) in terms of a state $\psi_N^{(0)}$ on the \ca\  $\CB_N$.
 
 At $N=\infty$, ground states of the system with \ca\ $\CB^l_{\infty}$ of quasi-local observables and  time-evolution $\ta$, see \er{brresult}, may be defined as  pure states $\om$ on $\CB^l_{\infty}$  
such that $-i\om(a^*\dl(a))\geq 0$ for each $a\in  \CB^l_{\infty}$ for which $\dl(a)=\lim_{t\raw 0} ((\ta_t(a)-a)/t)$ exists.\footnote{This the shortest  among many equivalent definitions: see e.g.\ Bratteli \&\  Robinson (1997, Definition 5.3.18) or Koma \& Tasaki (1994, App.\ A). For finite systems this (nontrivially, see refs.) reproduces the conventional definition of a ground state, and for infinite systems all known examples support its validity.}
 \item For $N<\infty$, ground states of the quantum Curie--Weisz model \er{CW} are defined as in the previous case, whereas at $N=\infty$ a ground state of the classical Hamiltonian \er{hcwc} is  a point $\vec{x}_0$ in the `phase space'  $\mathsf{B}^3$ that minimizes  $h^{CW}$  absolutely. 
 \end{enumerate}
 We now combine these definitions with the notion of a limit of a family of states (in the algebraic sense) following from our discussion of continuous fields in the previous chapter. That is, if $\mathfrak{A}^{\al}$ is our continuous field of \ca s, where $\al=c,j,g$ (see \S\ref{cfcas}),
and $(\om_x)_{x\in I_{\al}}$ is a family of states on this field in that $\om_x$ is a state on $\mathfrak{A}^{\al}_x$, then we say that 
\beq
\om_0=\lim_{x\raw 0} \om_x
\eeq
 if the $(\om_x)$ form a continuous field of states. For $\al=c$ this reproduces the notion of convergence $\rh_{\hbar}\raw\rh_0$ discussed after \er{defmuh}, whereas for $\al=g,l$ this gives meaning to limits like $\lim_{N\raw\infty}\om_{1/N}=\om_0$, where each $\om_{1/N}$ is a state on $\CB_N$ and $\om_0$ is a state on  $\mathfrak{A}_0^{g,l}$. 
 \begin{enumerate}
\item The quantum double-well Hamiltonian \er{TheHam2} has a unique ground state $\Psi_{\hbar}^{(0)}$, which (by a suitable choice of phase) may be chosen to be real and strictly positive.\footnote{See Reed \& Simon (1978, \S{XIII.12}). Uniqueness follows from an infinite-dimensional version of the Perron--Frobenius Theorem of linear algebra, which also yields strict positivity of the wave-function.}
Since the ground state is unique, it is  $\Z_2$-invariant (for otherwise its image under $u$ in \er{uz2} would be another ground state).
Seen as a wave-function,
$\Psi_{\hbar}^{(0)}$ has  well-separated peaks above $a$ and $-a$, which, as $\hbar\raw 0$, become increasingly pronounced.\footnote{See Landsman \& Reuvers (2013) for more details as well as some pictures.}

However, the classical double-well Hamiltonian \er{Hcl}  has \emph{two} ground states
\beq
\ps_0^{\pm}=(p=0,q=\pm a), \label{clgrstates}
\eeq 
which are  mapped into each other by the $\Z_2$-symmetry $(p,q)\mapsto (p,-q)$.
These states may be visualized as the particle being at rest at location $q=\pm a$.
Now define
\begin{equation}
\psi_0^{(0)}=\half(\psi_0^++\psi_0^-), \label{Cat1}
\end{equation}
where we identify $\psi_0^{\pm}\in\R^2$ with the corresponding (pure) state on $C_0(\R^2)$,
 so that the right-hand side is a convex sum of states (and hence  a state itself).\footnote{If in turn we identify  states 
on $C_0(\R^2)$ with probability measures on $\R^2$, as in \er{RMT}, then the right-hand side of \er{Cat1}
is a convex sum of probability measures. Specifically, we have
  $\psi_0^{(0)}(f)=\half(f(0,a)+f(0,-a))$.} Then the family $(\psi_{\hbar}^{(0)})_{\hbar\in[0,1]}$ forms a 
  a continuous field of states on the continuous field of \ca s $\mathfrak{A}^c$, and we have  the central result announced in \S\ref{savingp}, namely
  \begin{equation}
\lim_{\hbar\raw 0} \psi_{\hbar}^{(0)}=\psi_0^{(0)}.
\end{equation}
\item 
Similarly, for any $N<\infty$ and any $B>0$ the ground state $\Psi_N^{(0)}$ of the quantum Ising model \er{qising} is unique and hence $\Z_2$-invariant.\footnote{This was first established 
in Pfeuty (1970) by explicit calculation, based on Lieb et al (1961). This calculation, which is based on a Jordan--Wigner transformation to a fermonic model, cannot be generalized to higher dimensions $d$, but uniqueness of the ground state holds in any $d$, as first shown by
Campanino et al (1991) on the basis of Perron--Frobenius type arguments similar to those for  Schr\"{o}dinger operators. The singular case $B=0$ leads to a violation of the strict positivity conditions necessary to apply the 
 Perron--Frobenius Theorem, and this case  indeed features a degenerate ground state even when $N<\infty$.} The corresponding model at $N=\infty$ with small magnetic field $0\leq B<1$, on the other hand, has a doubly degenerate ground state $\ps_{\infty}^{\pm}$, in which  all spins are either up (+) or down (-).\footnote{See Araki \& Matsui (1985). For $B\geq 1$ all spins align in the $x$-direction and the ground state is unique.}
 If we relabel  the algebraic state corresponding to  $\Psi_N^{(0)}$ as  $\psi_{1/N}^{(0)}$, and similarly relabel
 $\ps_{\infty}^{\pm}$ as $\ps_0^{\pm}$, with corresponding mixture \er{Cat1}, then
 the states $(\ps_x)_{x\in I_l}$ form a continuous field of states on the continuous field of \ca s 
 $\mathfrak{A}^l$, and,\footnote{This a reformulation in our continuous field language of Corollary B.2 in Koma \& Tasaki (1994).}
   \begin{equation}
\lim_{N\raw\infty} \psi_{1/N}^{(0)}=\psi_0^{(0)}. \label{limitN0}
\end{equation}
\item The situation for the quantum Curie--Weisz model is the same, \emph{mutatis mutandis}:\footnote{See Rieckers (1981) and Gerisch (1993) for the analogue of \er{limitN0} in the quantum Curie--Weisz model. } for $N<\infty$ 
its ground state is unique and hence $\Z_2$-invariant,\footnote{This seems well known, but the first rigorous proof we are aware of is very recent (Ioffe \& Levit, 2013).} but for $N=\infty$ and  $0\leq B<1$ the classical Hamiltonian \er{hcwc} has two distinct ground states $\psi_0^{\pm}$, duly related by $\Z_2$ (here realized by a 180-degree rotation around the $x$-axis, 
cf.\ \S\ref{QICSec}).\footnote{As points of $\mathsf{B}^3$, for $0\leq B<1$ these are given by $\psi_0^{\pm}=(B,0,\pm\sin(\arccos(B)))$. For $B>1$ the unique ground state is $(1,0,0)$, which for $B=1$ is a saddle point. These points all lie on the boundary $S^2$ of $\mathsf{B}^3$.}
 \end{enumerate}
\section{First excited states}\label{TL}
In all three cases, the higher-level theories (i.e., classical mechanics for $\hbar=0$,   local quantum statistical mechanics at $N=\infty$, and classical thermodynamics) display \ssb\ of the $\Z_2$-symmetry of the Hamiltonian, whereas the corresponding  lower-level theories (that is, quantum mechanics and twice  quantum statistical mechanics at $N<\infty$) do not. Consequently, the ($\Z_2$-invariant) ground state of the lower-level theory in question cannot possibly converge to the ground state of the corresponding higher-level theory, because the latter fails to be  $\Z_2$-invariant (and the limiting process preserves $\Z_2$-invariance). 
Instead, one has \er{Cat1}, showing that the `lower-level' ground state converges to the `Schr\"{o}dinger Cat' state \er{Cat1}. This leads to the problems discussed at length in the Introduction. 

As already mentioned, the solution is to take the first excited state $\Psi_{\bullet}^{(1)}$ into account. 
\begin{enumerate}
\item For the double-well potential, the eigenvalue splitting 
$\Dl_{\hbar}\equiv E_{\hbar}^{(1)}-E_{\hbar}^{(0)}$ for small $\hbar$
is well known, see the heuristics in  Landau \& Lifshitz (1977), backed up by rigor in Simon (1985) and Hislop \& Sigal (1996).
The leading term as $\hbar\raw 0$ is
\beq
\Dl_{\hbar}\cong\frac{\hbar\om}{\sqrt{\half e\pi}}\cdot e^{-d_V/\hbar} \:\: (\hbar\raw 0), \label{ED}
\eeq
where the coefficient in the exponential decay in $-1/\hbar$ is the
  {\sc wkb}-factor
\beq
d_V=\int_{-a}^a dx\, \sqrt{V(x)}. \label{Cwkb}
\eeq
 On the basis of rigorous asymptotic estimates in Harrell (1980) and Simon (1985), which were subsequently verified by numerical computations,  Landsman \&\ Reuvers (2013) showed that the algebraic states $\ps^{\pm}_{\hbar}$ defined by the linear combinations
\begin{equation}
\Ps_{\hbar}^{\pm}= \frac{\Ps_{\hbar}^{(0)}\pm\Ps_{\hbar}^{(1)}}{\sqrt{2}}, \label{plusmin}
\end{equation}
satisfy
\beq
\lim_{\hbar\raw 0} \ps^{\pm}_{\hbar}=\ps_0^{\pm}.\label{1.7}
\eeq
\item The eigenvalue splitting for the quantum Ising chain (with free boundary conditions) can be determined on the basis of its exact solution by Pfeuty (1970). For the leading term in $\Dl_N\equiv  E_N^{(1)}-E_N^{(0)}$ as $N\raw \infty$, for $0<B<1$, we obtain\footnote{The first steps in this calculation are given by Karevski (2006), to which we refer for notation and  details.  To complete it, one has to solve his (1.51) for $v$  also to subleading order as $N\raw\infty$, noting first that his leading order solution $v=\ln(h)$ (where his $h$ is our $B$) has the wrong sign. To subleading order we find $v=-\ln(B) -(1-B^2)B^{2(N-1)}$. Subsituting $q_0=\pi+iv$ in the expression for the single-fermion
  excitation energy $E_N^{(1)}=\varep(q_0)=\sqrt{1+B^2+2B\cos(q_0)}$, one finds $\varep(q_0)=0$ to leading order and
  $\varep(q_0)=(1-B^2)B^N$ to subleasing order. But this is precisely $\Dl_N$, since in the picture of  Lieb et al (1961) the ground state is the fermonic Fock space vacuum, which has $E_N^{(0)}=0$. The energy splitting in higher dimensions does not seem to be known, but Koma \& Tasaki (1994, eq.\ (1.5)) expect similar behaviour.} 
\begin{equation}
\Dl_N\cong(1-B^2)B^N\:\: (N\raw\infty),
\end{equation}
showing exponential decay $\Dl_N\sim\exp(-CN)$ with positive coefficient $C=-\ln(B)$. 

\noindent Furthermore, the analogue of \er{1.7} holds (Koma \& Tasaki, 1994, App.\ B),\footnote{See especially  their Corollary B.2 and subsequent remark. This corollary is formulated in terms of the state
$\Ps_N=O_N\Ps_N^{(0)}/\|O_N\Ps_N^{(0)}\|$, where for the quantum Ising model one has $O_N=\sum_{i\in\underline{N}}\sg_i^z$, but since $\|\Ps_N^{(1)}-\Psi_N\|=O(1/N)$ as $N\raw\infty$ by their Lemma B.4, our \er{1.7b} follows. A similar comment applies to the quantum Curie--Weisz model (taking into account the subtlety to be mentioned in the next footnote).} viz.
\beq
\lim_{N\raw\infty} \ps^{\pm}_{N}=\ps_{\infty}^{\pm},\label{1.7b}
\eeq
where $\ps^{\pm}_{N}$ are the algebraic states defined by the unit vectors
\begin{equation}
\Ps_N^{\pm}= \frac{\Ps_N^{(0)}\pm\Ps_N^{(1)}}{\sqrt{2}}. \label{plusmin2}
\end{equation}
\item For the quantum Curie--Weisz model, exponential decay of $\Dl_N$ has only been established
numerically up to $N\sim 150$ (Botet, Julien \& Pfeuty, 1982; Botet \& Julien, 1982).\footnote{Se also  related simulations up to $N=1000$ in Vidal et al (2004). In the worst case, where for whatever reasons these simulations are misleading,  $\Dl_N$ would decay as $1/N$, which does not jeopardize our scenario, but would add some constraints on the perturbations destabilizing the ground state, see below. This decay follows from Theorem 2.2 in Koma \& Tasaki (1994), where it has to be noted that one of the assumptions in their proof (namely that the support set of $H_i$, where $H_N=\sum_i H_i$, is bounded in $N$) is invalid  in the quantum Curie--Weisz model;
nonetheless, their eq.\ (2.11) can be proved by direct computation. } Eq.\  \er{1.7b} may be proved in the same way as in the quantum Ising model.\footnote{It would be interesting to apply the techniques of 
Ioffe \& Levit (2013), designed for  $\Ps_N^{(0)}$, to $\Ps_N^{(1)}$.}\end{enumerate}
In summary, in each of our models we may define the unit vectors
\begin{equation}
\Ps_x^{\pm}= \frac{\Ps_x^{(0)}\pm\Ps_x^{(1)}}{\sqrt{2}}, \label{plusmin2}
\end{equation}
where  $x=\hbar$ or $x=1/N$. Given the exponential decay of the eigenvalue splitting,
in the asymptotic regime $\hbar\raw 0$ or $N\raw\infty$ these are `almost' energy eigenstates. Indeed, the corresponding algebraic states  converge to time-independent states of the  limit theories:
\beq
\lim_{x\raw 0} \ps^{\pm}_x=\ps_0^{\pm}.\label{1.7bis}
\eeq
As explained in the Introduction, this would remove the asymptotic emergence paradox, and hence rescue Earman's and Butterfield's Principles, provided that two conditions hold:
\begin{enumerate}
\item The \emph{degenerate} ground states $\ps_0^+$ or $\ps_0^-$ of the \emph{higher-level}  theory $\mathsf{H}$ are \emph{stable}, so that for  small $x$ any  approximant to either within $\mathsf{L}_x$  is at least approximately stable. 
\item The \emph{unique} ground states $\Ps_x^{(0)}$ of the \emph{lower-level} theories $\mathsf{L}_x$, on the other hand,  become \emph{unstable} as
$x$ decreases; when perturbed within $\mathsf{L}_x$, they move to states $\Ps_x'$ that, as $x\raw 0$, converge to either one of the pure states $\ps_0^+$ or $\ps_0^-$ of  $\mathsf{H}$;
\end{enumerate}

 The first point is standard, see e.g. Bratteli \&\ Robinson (1997, p.\ 174) for large systems, and any book on classical mechanics for our first example (intuitively, small kicks to a ball at the bottom of a potential well leave the ball in that well).

The considerably more surprising second point was established for our first model, i.e., the double-well potential, in the fundamental work of Jona-Lasinio, Martinelli, \& Scoppola (1981a,b); see also Helffer \&\ Sj\"{o}strand (1986) and Simon (1985). 

This work was applied to the \mmp\ in Landsman \&\ Reuvers (2013); as explained in the Introduction, their conclusions equally well apply to \ssb.\footnote{Indeed, using the `interaction matrix' formalism of Helffer \&\ Sj\"{o}strand (1986) and Simon (1985), an analogous analysis can be given for the quantum statistical mechanics models. This works because in these models (in which the energy spectrum is discrete for any  $N<\infty$), as $N\raw\infty$, 
   on the one hand all energy levels merge into a continuum, but
 on the other, they split into pairs whose energy difference is exponentially small.  In addition,  the instability of the ground state has been derived in a different way by
 Narnhofer  \&\ Thirring (1996, 1999) for $\Z_2$, as well as by van Wezel (2007, 2008, 2010) for $SU(2)$.}
Our scenario requires the system to be coupled to something like an environment (which may also be internal to the system at hand),\footnote{As explained in Landsman \& Reuvers (2013) in connection with the measurement problem, this is \emph{not} the decoherence scenario, which unlike ours fails to predict individual 
measurement outcomes and hence fails to solve the the measurement problem  (it rather reconfirms it). In our present context of \ssb, mere decoherence would achieve nothing either, leading to the mixtures $\psi_0^{(0)}$ rather than the pure states $\psi_0^{\pm}$.} but the point is that the  perturbations that are required may be almost arbitrarily small: whatever their size, as long as they are asymmetric, they do their destabilizing work increasingly well as $\hbar\raw 0$ or $N\raw\infty$. 

To get a quick (though potentially misleading) feeling for what is going on here, as in Landsman \&\ Reuvers (2013), take a basis of the two-level \Hs\ $\C^2$ consisting of
$e_1=\Psi_x^+$ and $e_2=\Psi_x^-$ (as opposed to $\Ps_x^{(0)}$ and $\Ps_x^{(1)}$). 
We write  $\Dl_x$ for the pertinent eigenvalue splitting, where $x$ is $\hbar$ or $1/N$, and $\Dl_{1/N}\equiv\Dl_N$, and define the matrix 
 \begin{equation}
H^{(\dl)}_x=\half
\left(
\begin{array}{cc}
 \dl_+ & -\Dl_x   \\
 -\Dl_x &  \dl_-
 \end{array}
\right). \label{HD2}
 \end{equation}
For $\dl_{\pm}=0$, $H_x^{(0)}$ acts like an effective Hamiltonian that correctly  reproduces the (unperturbed) ground state
$(e_1+e_2)/\sqrt{2}=\Ps_x^{(0)}$. Now add perturbations $\dl_{\pm}$.
Since  $\Dl_x$ vanishes exponentially quickly as $x\raw 0$, almost any nonzero choice of the $\dl_{\pm}$  has the effect that as $x\raw 0$, the perturbed Hamiltonian \er{HD2} is dominated by its diagonal, as opposed to the original  Hamiltonian $H_x^{(0)}$. Hence with the exception of the unlikely symmetric case $\dl_+=\dl_-$, or the case where the $\dl_{\pm}$ vanish as quickly as $\Dl_x$ as $x\raw 0$, 
 the ground state of $H^{(\dl)}_x$ will tend to either $e_1$ or $e_2$ as $x\raw 0$. Consequently, the effect of arbitrary asymmetric perturbations is to change the ground state of the system from a Schr\"{o}dinger Cat one like $\Ps_x^{(0)}$ to a localized one like $\Ps_x^{\pm}$ (either in physical space, as for the double well, or in spin configuration space, as in our other two models).\footnote{ The interaction matrix formalism makes this reasoning rigorous. As in footnote \ref{fwarning}, we stress that the quick argument given here is only meant to illustrate the \emph{instability} of the ground state $\Ps_x^{(0)}$in $\mathsf{L}_x$; it gives the wrong picture of the asymptotic \emph{stability} of the states $\Ps_x^{\pm}$. In particular, in the actual interaction matrix formalism as presented in Simon (1985) the basis of $\C^2$ is not given by the vectors $\Psi_x^{\pm}$ but by the projections of the solutions of the cutoff Schr\"{o}dinger equation (with double-well potential) localized in $\R^{\pm}$ with Dirichlet boundary conditions at 0 onto the linear subspace spanned by the lowest two eigenstates of the \emph{perturbed} Hamiltonian. }
\medskip
 
What remains to be done theoretically is to first model the perturbations achieving this dynamically (i.e., in time), and subsequently to study also the dynamical transition from the original, delocalized, unperturbed ground state to the perturbed, localized  ground state. For the double-well case this program has been started in Landsman \& Reuvers (2013),  and for the spin systems this is a matter for future research.

Experimentally, the entire scenario should be put to the test (which will not be easy, given the discrepancy between what is considered large $N$ or small $\hbar$ in theory and in laboratory practice);  this will be done in the near future in the context of a larger  project \emph{Experimental Tests of Quantum Reality},  led by Andrew Briggs at Oxford.
\newpage
\section*{References}
\begin{small}
\begin{trivlist}
\item  Allahverdyana, A.E.,  Balian, R., \&\  Nieuwenhuizen, Th.M. (2013).
Understanding quantum measurement from the solution of dynamical models. \emph{Physics Reports}
525, 1--166. 
\item  Anderson, P.W. (1952). An approximate quantum theory of the antiferromagnetic ground state.
  \emph{Physical Review} 86, 694--701.
\item Anderson, P.W. (1972). More is different. \emph{Science} 177, 4047, 393--396.
\item Anderson, P.W. (1984). \emph{Basic Notions of Condensed Matter Physics}. Boulder: Westview Press. 
\item Araki, H.,\&  Matsui, T. (1985). Ground states of the $XY$-model. \emph{Communications in Mathematical Physics} 101, 213--245.
\item Batterman, R. (2002). \emph{The Devil in the Details: Asymptotic Reasoning in Explanation, Reduction, and Emergence}. Oxford: Oxord University Press.
\item Batterman, R. (2005). Response to Belot's ``Whose devil? Which details?''.
 \emph{Philosophy of Science} 72, 154--163.
\item Batterman, R. (2011). Emergence, singularities, and symmetry breaking. \emph{Foundations of Physics}
41, 1031--1050. 
\item Belot, G. (2005). Whose devil? Which details? \emph{Philosophy of Science} 72, 128--153.
\item Bona, P. (1988). The dynamics of a class of mean-field theories. \emph{Journal of Mathematical Physics} 29, 2223--2235.
\item  Botet, R., \& Julien, R. (1982). Large-size critical behavior of  infinitely coordinated systems. \emph{Physical Review} B28, 3955--3967.
\item  Botet, R.,  Julien, R., \& Pfeuty, P. (1982).
Size scaling for infinitely coordinated systems. \emph{Physical Review Letters}
49, 478--481.
\item  Bratteli, O., \&\  Robinson, D.W. (1997). \emph{
Operator Algebras and Quantum Statistical Mechanics. Vol.\ II:
Equilibrium States, Models in Statistical Mechanics}, 2nd ed. Berlin: Springer.
\item Broad, C.D. (1925). \emph{The Mind and its Place in Nature}. London: Routledge \& Kegan Paul.
\item Butterfield, J. (2011). Less is different: Emergence and reduction reconciled.
\emph{Foundations of Physics} 41, 1065--1135.
\item Butterfield, J., \& Bouatta, N. (2011). Emergence and reduction combined in phase transitions.
\emph{Proc.\ Frontiers of Fundamental Physics (FFP11)}.  
\item  Campanino, M.,  Klein, A., \& Perez, J.F. (1991).  Localization in the ground state of the Ising model with a random transverse field. \emph{Communications in Mathematical Physics} 135, 499-515.
\item  Dixmier, J. (1977). \emph{
$C^*$-Algebras}. Amsterdam: North-Holland.
\item Duffield, N.G., \& Werner, R.F. (1992). Local dynamics of mean-field quantum systems. \emph{Helvetica Physica Acta} 65, 1016--1054. 
\item Earman, J. (2003). Rough guide to spontaneous symmetry breaking. \emph{Symmetries in Physics: Philosophical Reflections}, eds. Brading, K., \&  Castellani, E., pp. 335--346. Cambridge: Cambridge 
University Press. 
\item Earman, J. (2004). Curie's Principle and spontaneous symmetry breaking. \emph{International Studies in the Philosophy of Science} 18, 173--198. 
\item Gerisch, T. (1993). Internal symmetries and limiting Gibbs states in quantum lattice mean-field models.
\emph{Physica} A197, 284--300. 
\item   Haag, R. (1992).   \emph{Local Quantum Physics: Fields, Particles, Algebras}. 
Heidelberg: Springer.  
\item  Harrell, E.W. (1980). Double wells.
\emph{Communications in Mathematical Physics} 75, 239--261. 
\item Helffer, B. (1988). \emph{Semi-classical Analysis for the Schr\"{o}dinger Operator and Applications (Lecture Notes in Mathematics 1336)}.
Heidelberg: Springer.
\item Helffer,  B., \&  Sj\"{o}strand,  J. (1986).
  R\'{e}sonances en limite semi-classique.
   \emph{M\'{e}moires de la Soci\'{e}t\'{e} Math\'{e}matique de France (N.S.)}  24--25, 1--228.
   \item Hempel, C., Oppenheim, P., (2008) [1965]. On the idea of emergence.
 \emph{Emergence}, eds.\ Bedau, M.A., \& Humphreys, P., pp.\ 61--80. Cambridge (Mass.): {\sc mit} Press.
 Originally published in Hempel, C., \emph{Aspects of Scientific Explanation and Other Essays in the Philosophy of Science}. 	
New York: The Free Press. 
\item  Hislop, P.D.,  \& Sigal,   I.M. (1996). \emph{Introduction to Spectral Theory}. New York:
Springer.
\item Hooker, C.A. (2004). Asymptotics, reduction and emergence. \emph{British Journal for the Philosophy of Science} 55, 435--479.
\item Ioffe, D., \& Levit, A. (2013). Ground states for mean field models with a transverse
component. \emph{Journal of Statistical Physics}  151, 1140--1161.
\item  Jona-Lasinio, G., Martinelli, F.,  \& Scoppola, E. (1981a).
New approach to the semiclassical limit of quantum mechanics.
\emph{Communications in Mathematical Physics}  80, 223--254.
\item 
Jona-Lasinio, G., Martinelli, F.,   Scoppola, E. (1981b).
 The semiclassical limit of quantum mechanics: A qualitative theory via stochastic mechanics.  \emph{Physics Reports} 77, 313--327.
\item Jones, N.J. (2006). \emph{Ineliminable Idealizations, Phase Transitions, and Irreversibility}.
PhD Thesis, Ohio State University. 
\item Kadison, R.V., \&\ Ringrose, J.R. (1986). \emph{Fundamentals of the Theory of Operator Algebras. Vol. 2: Advanced  Theory}.   New York: Academic Press.
\item Karevski, D. (2006). \emph{Ising Quantum Chains}. \texttt{arXiv:hal-00113500}.
\item Kirchberg, E., \&\
  Wassermann, S. (1995). Operations on continuous bundles of
 $C^*$-algebras. \emph{Mathematische  Annalen}  303, 677--697.
\item Koma, T.  and  Tasaki, H.  (1994). Symmetry breaking and finite-size effects in quantum many-body systems.
\emph{Journal of Statistical Physics} 76, 745--803.
\item Landsman, N.P. (1998). \emph{Mathematical Topics Between Classical and Quantum Mechanics.}
New York: Springer.
\itemÊLandsman, N.P.  (2007). Between classical and quantum. \emph{Handbook of the Philosophy of Science Vol. 2: Philosophy of Physics, Part A}, eds. Butterfield, J., \& Earman, J., 
pp. 417-553. Amsterdam: North-Holland.
\itemÊLandsman, N.P., \& Reuvers, R. (2013). A flea on Schr\"{o}dinger's Cat.
 \emph{Foundations of Physics} 43, 373--407.
\item Landau, L.D., \&  Lifshitz, E.M. (1977). \emph{Quantum Mechanics}, Third Ed. Oxford: Pergamon Press.
\item Lieb, E., Schultz, T., \& Mattis, D. (1961). Two soluble models of an antiferromagnetic chain.
\emph{Annals of Physics} 16, 407--466.
\item  Liu, C.,  \&  Emch, G.G.  (2005). Explaining quantum spontaneous symmetry breaking.
\emph{Studies In History and Philosophy of Modern Physics}  36, 137--163. 
 \item Marsden, J.E. \&\ T.S. Ratiu (1994).
\emph{Introduction to Mechanics and Symmetry}. New York: Springer.
\item  McLaughlin, B.P. (2008). The rise and fall of British Emergentism. \emph{Emergence}, eds.\ Bedau, M.A., \& Humphreys, P., pp.\ 19--59. Cambridge (Mass.): {\sc mit} Press.
\item Menon, T., \& Callender, C. (2013). Turn and face the strange\ldots Ch-ch-changes: Philosophical questions raised by phase transitions. \emph{The Oxford Handbook of Philosophy of Physics}, ed.  Batterman, R., 
pp.\ 189--223.
New York:  Oxford University Press.
\item Mill, J.S. (1952 [1843]). \emph{A System of Logic, 8th ed}. London: Longmans, Green, Reader, and Dyer.
\item Narnhofer, H., \&\ Thirring, W. (1996). Why Schr\"{o}dinger's Cat is most likely to be either alive or dead.
Preprint ESI 343. Available at \texttt{http://www.mat.univie.ac.at/$\sim$esiprpr/esi343.pdf}.
\item Narnhofer, H., \&\ Thirring, W. (1999). Macroscopic purification of states by interactions.
\emph{On Quanta, Mind and Matter: Hans Primas in Context}, eds.\ Atmanspacher, H., et al, 
 pp.\ 105--118. Dordrecht: Springer. 
\item Norton, J.D. (2012). Approximation and idealization: Why the difference matters.
\emph{Philosophy of Science} 79, 207--232.
\item
O'Connor, T. \& Wong, H.Y. (2012). Emergent Properties. \emph{The Stanford Encyclopedia of Philosophy (Spring 2012 Edition)}, ed.  Zalta, E.N.
\item Pfeuty,  P. (1970). The one-dimensional {I}sing model with a transverse field. \emph{Annals of Physics}
47, 79--90.
\item Paul, T., \&  Uribe, A. (1996). On the pointwise behavior of semi-classical measures.  
 \emph{Communications in Mathematical Physics}
  175, 229--258.
   \item Raggio, G.A., \&\ Werner, R.F. (1989). Quantum statistical mechanics
  of general mean field systems. \emph{Helvetica Physica Acta} 62, 980--1003.
\item Reed, M. \&\ Simon, B. (1978).  \emph{Methods of Modern Mathematical Physics. Vol IV. Analysis of Operators.} New York: Academic Press.
\item Rieckers, A. (1981). Effective dynamics of the quantum mechanical Wei\ss-Ising model. \emph{Physica} A108, 107--134. 
\item   Robert, D. (1987).  \emph{Autour de l'Approximation
 Semi-Classique}. Basel: Birkh\"{a}user. 
  \item Rueger, A. (2000). Physical emergence, diachronic and synchronic. \emph{Synthese} 124, 297--322.
 \item Rueger, A. (2006). Functional reduction and emergence in the physical sciences. \emph{Synthese} 151, 335--346.
 \item Ruetsche, L. (2011). \emph{Interpreting Quantum Theories.} Oxford: Oxford University Press.
\item Sachdev, S. (2011). \emph{Quantum Phase Transitions}, 2nd ed. Cambridge: Cambridge University Press.
\item Silberstein, M. (2002). Reduction, emergence and explanation.  \emph{The Blackwell Guide to the Philosophy of Science}, eds. Machamer, P.K., \& Silberstein, M., pp.\ 80--107.
Oxford: Blackwell. 
\item  Simon, B. (1985). Semiclassical analysis of low lying eigenvalues. IV. The flea on the elephant,  \emph{J. Funct. Anal.} 63, 123--136.
\item Stephan, A. (1992). Emergence - a systematic view of its historical facets.
 \emph{Emergence or Reduction?}, ed.\  Beckermann, A., pp.\ 25--48. Berlin: De Gruyter.
\item Stein, E.M, \& Shakarchi, R. (2005). \emph{Real Analysis: Measure Theory, Integration, and Hilbert Spaces}.
Princeton: Princeton University Press.  
\item St\o rmer, E. (1969). Symmetric states of infinite tensor products of \ca s. \emph{Jornal of Functional Analysis} 3, 48--68.
\item Suzuki, S., Inoue, J.-i., \& Chakrabarti, B.K. (2013). \emph{Quantum Ising Phases and Transitions in Transverse Ising Models}, 2nd ed. Heidelberg: Springer. 
\item Vidal, J., Palacios, G., \& Mosseri, R. (2004). Entanglement in a second-order quantum phase transition.
\emph{Physical Review} A69, 022107. 
\item  Wayne, A., \& Arciszewski, M. (2009). Emergence in physics. \emph{Philosophy Compass} 4/5, 846--858.
\item Wezel, J. van (2007). \emph{Quantum Mechanics and the Big World}. PhD Thesis, Leiden University, available at
\texttt{https://openaccess.leidenuniv.nl/handle/1887/11468}. 
\item Wezel, J. van (2008). Quantum dynamics in the thermodynamic limit. \emph{Physical Review B} 78, 054301.
\item Wezel, J. van (2010). Broken time translation symmetry as a model for quantum state reduction.
\emph{Symmetry}  2, 582--608.
\item Wezel, J. van, Brink, J. van den, \&\ Zaanen, J. (2005). 
An intrinsic limit to quantum coherence due to spontaneous symmetry breaking. \emph{Physical Review Letters}
 94, 230401.
\end{trivlist}
\end{small}
\end{document}